\documentclass[prl,twocolumn,nofootinbib]{revtex4-2}

\usepackage{pifont} 

\usepackage[utf8]{inputenc}
\usepackage[a4paper]{geometry}
\usepackage{hyperref}
\usepackage{graphicx}
\usepackage{xcolor}
\usepackage{multirow}
\usepackage[normalem]{ulem}

\usepackage{setspace}
\usepackage[caption=false]{subfig}
\usepackage[T1]{fontenc}

\usepackage{amsmath}
\usepackage{amssymb}
\usepackage{amsfonts, relsize, color}
\usepackage{mathrsfs}
\usepackage{mathtools}
\usepackage{physics}
\usepackage{bbm}
\usepackage{bbold}

\newcommand{\ncmd}{\newcommand}
\ncmd{\nn}{\nonumber}
\ncmd{\mbf}[1]{\bs{#1}}
\ncmd{\gam}{\gamma}
\ncmd{\sig}{\sigma}
\ncmd{\pha}{\alpha}
\ncmd{\lam}{\lambda}
\ncmd{\kap}{\kappa}
\ncmd{\Lam}{\Lambda}
\ncmd{\Gam}{\Gamma}
\ncmd{\Ups}{\Upsilon}
\ncmd{\Om}{\Omega}
\ncmd{\eps}{\epsilon}
\ncmd{\veps}{\varepsilon}
\ncmd{\vphi}{\varphi}
\ncmd{\vtheta}{\vartheta}
\ncmd{\tw}{\text{w}}
\ncmd{\pd}{\partial}
\ncmd{\pll}{\parallel}
\ncmd{\mc}{\mathcal}
\ncmd{\mf}{\mathfrak}
\ncmd{\dl}{\delta}
\ncmd{\bs}{\vec}
\ncmd{\half}{\frac{1}{2}}
\ncmd{\tilJ}{\widetilde{J}}
\ncmd{\avg}[1]{\langle{#1}\rangle}
\ncmd{\note}[1]{{\color{red}{\ding{168} [#1]}}}
\ncmd{\eq}[1]{Eq. \eqref{#1}}
\ncmd{\fig}[1]{Fig. \ref{#1}}
\ncmd{\suppl}{\note{`Supplementary Information'}}
\ncmd{\kl}[1]{{\color{blue}#1}}

\begin{document}

\title{Topological anisotropic non-Fermi liquid from a Berry-dipole semimetal}

\author{Konstantinos Ladovrechis}
\email{konstantinos.ladovrechis@ruhr-uni-bochum.de}
\affiliation{ 
	Theoretische Physik III, Ruhr-Universität Bochum, D-44801 Bochum, Germany}	
	
	\begin{abstract}
	The interplay among topology and interactions has been a promising path towards identifying novel phases of condensed matter beyond these predicted by the established classification paradigms. In the present work, we propose such a novel phase of matter by studying the fate of a three-dimensional Berry-dipole semimetal, lying at the topological quantum critical point separating a Hopf insulator from a trivial insulator, in the presence of long-range Coulomb interactions. Utilizing large-$N_f$ analysis and an $\epsilon$-expansion within the renormalization-group scheme, we predict the emergence of a spatially anisotropic non-Fermi liquid with enhanced Berry-dipole moment. The corresponding scaling relations of certain physical observables are derived as functions of the probed energy and temperature scale, and we conclude by providing an observational test for probing the enhanced topological features of the anisotropic non-Fermi liquid.
		
	\end{abstract}
	
	\date{\today}
	
	\maketitle

	\textit{Introduction}.~The concepts of topology and interactions have been fundamental research disciplines for condensed-matter physics in the last three decades \cite{Hasan2010,Qi2011,Sachdev2011,Rachel2018}. In the presence of a finite Fermi surface, they might appear as synergistic or competing constituents as far as the \textit{full} spectrum and response functions are concerned. When interactions are ignored, the single-particle formalism allows for a classification of the corresponding phases of matter \cite{Turner2013,Chiu2016,Armitage2018}. However, in the general case, interactions cannot be ignored, and it is yet unclear and largely unknown how a systematic categorization of the phases of matter can be facilitated. In particular, long-range Coulomb interactions in topological semimetals lead to a vanishing quasi-particle residue and to the invalidation of a single-particle representation. The result is the emergence of interaction-induced non-Fermi liquids lying \textit{beyond} the predictive power of classification rules. Such phases of matter have been proposed to exist in Luttinger semimetals (see Ref.~\cite{Zhu2024} for the features of delicate topology), Dirac, and Weyl semimetals \cite{Savary2014,Boettcher2017,Lai2015,Jian2015,Cho2016,Yang2014} as well as at topological quantum critical points (TQCPs) distinguising double- and triple-Weyl semimetals from a trivial insulator \cite{Han2019,Wang2019}. A possible route towards utilizing correlations to generate novel phases of matter beyond the standard classification with potential underlying topology is the case of long-range Coulomb interactions on the TQCP separating the time-reversal-breaking topological \textit{Hopf insulator} from a trivial insulator \cite{Moore2008,Deng2013}. In the low-energy regime, such a TQCP is described in terms of a \textit{Berry-dipole semimetal}, a three-dimensional fermionic system with quadratic band-touching at the origin of the Brillouin zone supporting a finite Berry-dipole moment \cite{Nelson2022,Zhuang2024}. The question for the existence of novel phases of matter as derivatives of a Berry-dipole semimetal becomes even more relevant due to the recent experimental realization of the latter in acoustic lattices \cite{Mo2025}, on one hand, and of Hopf insulators in nitrogen-vacancy centers and topoelectric circuits \cite{Yuan2017,Wang2023}, on the other hand.
	\par In this Letter, we propose the existence of a novel correlation-driven topological phase of matter as derivative of the Berry-dipole semimetal. The vanishing of the density of states in the vicinity of the origin of the Brillouin zone causes the long-range Coulomb interaction to be a relevant perturbation from a renormalization-group point of view \cite{Herbut2007,Herbut2014}. The considered Berry-dipole semimetal manifests a four-fold rotational symmetry around the $\hat{z}$ axis, which indicates potential differentiation in the scaling of the form factors under Coulomb interactions. Within a large-$N_f$ analysis corroborated by an $\epsilon$-expansion through the renormalization-group scheme, we find that the interaction-induced semimetal shows strong spatial anisotropy resulting in scaling relations for physical observables significantly different from the Fermi-liquid estimates \cite{Lohneysen2007}. Finally, the resulting enhancement of the fluxes of Berry curvature through the upper- and lower-half $z$-planes enables us to provide an observational test for the onset of the topological anisotropic non-Fermi-liquid behavior in terms of the instrinsic and extrinsinc non-linear Hall conductivities.
	\par
\begin{figure}[t]
		\includegraphics[scale=1]{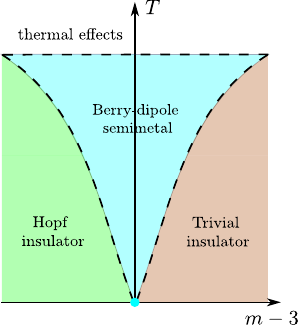}
		\caption{Schematic phase diagram of the non-interacting Hamiltonian matrix given in Eq.~\eqref{eq:HopfLattice} in the low-temperature regime in the vicinity of the TQCP $m=3$. The parabolic-shaped dashed lines denote the phase boundaries between the semimetallic and the two insulating phases, whereas the horizontal dashed line indicates the upper threshold beyond which thermal effects are no longer negligible.}
		\label{fig:PhaseDiagram}
	\end{figure}

 \textit{Model}.~We consider the following three-dimensional tight-binding model of Hopf insulators \cite{Moore2008}
	\begin{equation}\label{eq:HopfLattice}
		H(\bold{k}, m) = \bold{w}(\bold{k}, m) \cdot \boldsymbol{\sigma},
	\end{equation}
	where the three-dimensional momentum is denoted by vector $\bold{k}$. The vector of Pauli matrices is $\boldsymbol{\sigma}=(\sigma_1,\sigma_2,\sigma_3)$, and the vector $\bold{w}(\bold{k}, m)$ reads
	\begin{equation}\label{eq:ColumnVector}
		\bold{w}(\bold{k}, m) = \begin{bmatrix}
			2u_1(\bold{k})\,u_3(\bold{k}) + 2u_2(\bold{k})\,u_4(\bold{k}, m)\\
			2u_1(\bold{k})\,u_4(\bold{k}, m) - 2u_2(\bold{k})\,u_3(\bold{k})\\
			u_1^2(\bold{k}) + u_2^2(\bold{k}) - u_3^2(\bold{k}) - u_4^2(\bold{k}, m)
		\end{bmatrix},
	\end{equation}
	where $u_1(\bold{k})=t_p\sin{k_x}$, $u_2(\bold{k})=t_p\sin{k_y}$, $u_3(\bold{k})=t_z\sin{k_z}$, $u_4(\bold{k}, m)=\cos{k_x}+ \cos{k_y}+ \cos{k_z} - m$, and $\{t_p,t_z\}$ are positive-valued parameters. Four-fold rotational symmetry is present through the operation
	\begin{equation}\label{eq:C4z}
		\mc C_{4}^z:\,H(\bold{k},m)\,\, \mapsto\,\, e^{-i \frac{\pi}{4}\sigma_3} H^*(k_y, -k_x, -k_z,m) e^{i \frac{\pi}{4}\sigma_3}.
	\end{equation}
	Eq.~\eqref{eq:HopfLattice} also respects the anti-unitary mirror operations~\cite{Tyner2024}
	\begin{flalign}\label{eq:MxMy}
		&\mc M_x \mc T:\, H(\bold{k},m)\,\, \mapsto\,\, \sig_3 H^*(k_x, -k_y, -k_z,m)\sig_3,\nonumber\\
		&\mc M_y \mc T:\, H(\bold{k},m)\,\, \mapsto\,\, H^*(-k_x, k_y, -k_z,m),
	\end{flalign}
	where $\mc T$ implements spinless time reversal. The Hamiltonian in Eq.~\eqref{eq:HopfLattice} describes distinct Hopf-insulating and topologically trivial phases \cite{Moore2008,Deng2013}. In the present work, we focus on the TQCP at $m = 3$, where the two electronic bands touch quadratically at the origin of the Brillouin zone, i.e.~the $\Gamma$ point. In Fig.~\ref{fig:PhaseDiagram}, we schematically draw the expected non-interacting phase diagram in the vicinity of that TQCP in the low-temperature regime \cite{Vojta2003}. In the low-energy limit, that band-touching point acts as a simultaneous source and sink of Berry curvature exhibiting a quantized flux through the upper- and lower-half $z$-planes. Thus, it describes a Berry-dipole semimetal mediating a topological phase transition between a Hopf and a trivial insulator \cite{Nelson2022,Zhuang2024}. At the TQCP, the low-energy physics is predominantly controlled by momentum states close to the $\Gamma$ point. In this regard, we consider the effective single-particle Hamiltonian \cite{Ladovrechis2025} 
	\begin{equation}\label{eq:QuadraticHamiltonian}
		H_3(\bold{k}) = 2t^2_1k_z\qty[k_x\sigma_1+k_y\sigma_2]+\qty[t^2_2(k_x^2+ k_y^2)-t^2_3k^2_z]\sigma_3,
	\end{equation}
	where we have substituted expressions involving the tight-binding parameters $\{t_p,t_z\}$ in the form factors with three generic \textit{scale-dependent} anisotropy parameters $\{t^2_1,t^2_2,t^2_3\}$. We note that the Hamiltonian $H_3(\bold{k})$ preserves the symmetry operations described in Eqs.~\eqref{eq:C4z}--\eqref{eq:MxMy} while respecting a $z$-mirror symmetry as well \cite{Nelson2022}. Introducing the Grassmanian field
	$\psi^{(\dag)}(\bold{k},i\omega)$ as a two-dimensional Dirac spinor, where $\omega$ is the fermionic Matsubara frequency, the bare effective action in the presence of long-range Coulomb interactions reads
	\begin{flalign}\label{eq:ActionThreeDimensions}
		&\dfrac{\mathcal{S}}{T}=\sum_{i\omega}\int\dfrac{d^3k}{(2\pi)^3}\,\psi^\dag(\bold{k},i\omega)\left[-i\omega\mathbb{1}_2+H_3(\bold{k})\right]\psi(\bold{k},i\omega)\nonumber\\
		&+\sum_{i\Omega}\int\dfrac{d^3q}{(2\pi)^3}\phi^\dag(\bold{q},i\Omega)\dfrac{c^2_p(q^2_x+q^2_y)+c^2_zq^2_z}{2}\phi(\bold{q},i\Omega)\nonumber\\
		&-ie\sum_{i\omega}T\sum_{i\Omega}\int\dfrac{d^3k}{(2\pi)^3}\int\dfrac{d^3q}{(2\pi)^3}\,\psi^\dag(\bold{k}+\bold{q},i\omega+i\Omega)\times\nonumber\\
		&\times\phi(\bold{q},i\Omega)\psi(\bold{k},i\omega).
	\end{flalign}
	Parameter $\bold{q}$ $(\Omega)$ denotes the bosonic momentum (Matsubara frequency), and the bosonic field $\phi(\bold{q},i\Omega)$ is the corresponding Hubbard-Stratonovitch field \cite{Kopietz1997}. Parameter $e$ is the electronic charge, and parameters $\{c^2_p,c^2_z\}$ characterize the potentially different scaling for the bosonic form factors in the presence of Coulomb interactions. Due to the vanishing density of states in the zero-energy limit, the Coulomb potential is insufficiently screened by the electronic modes, thus the long-range Coulomb interaction is a relevant perturbation to the non-interacting Hamiltonian $H_3(\bold{k})$ \cite{Herbut2014}. To deal with such a perturbation, we employ first the large-$N_f$ method at three spatial dimensions to obtain a non-perturbative estimation for the renormalization pattern of the bare action in Eq.~\eqref{eq:ActionThreeDimensions} \cite{Ferrell1972,Abrikosov1974}.
	\par\textit{Large-$N_f$ method}.~We perform a large-$N_f$ analysis considering a \textit{frequency-dependent} bosonic field to ensure a well-behaved and physically consistent framework which avoids issues related to IR divergencies and the ``unique'' nature of the zero-frequency bosonic mode \cite{Wang2016,Wang2018,Schrodi2019}. Extending the dimension of the Dirac spinor $\psi^{(\dag)}(\bold{k},i\omega)$ from two to $2N_f$ components, where $N_f$ is the fermionic flavour number, the one-loop bosonic self-energy reads
	\begin{equation}\label{eq:LargeNfBosonicSelfEnergy}
		\dfrac{\Pi(\boldsymbol{q},i\Omega)}{N_fe^2T}=\sum_{i\omega}\int\dfrac{d^3k}{(2\pi)^3}\Tr G_3(\bold{k}+\bold{q},i\omega+i\Omega)G_3(\bold{k},i\omega),
	\end{equation}
	where we denote the non-interacting fermionic Matsubara Green's function as $G_3(\bold{k},i\omega)=\left[-i\omega\mathbb{1}_2+H_3(\bold{k})\right]^{-1}$. Upon performing summation over the fermionic Matsubara frequencies $\omega$ in Eq.~\eqref{eq:LargeNfBosonicSelfEnergy},
	we find that the bosonic self-energy is adequately described by the ansatz
	\begin{flalign}\label{eq:LargeNfBosonicSelfEnergyAnsatz}
		&\Pi(\boldsymbol{q},i\Omega)=-\dfrac{N_fe^2}{(2\pi)^3t^2_1}\sqrt{a^2_p(\beta)q^2_p+a^2_z(\beta)\beta^4_{31}q^2_z}\,\times\nonumber\\
		&\times\tanh(\sqrt{b^2_p(\beta)\dfrac{t^2_1}{\beta^2_{31}}\dfrac{q^2_p}{|\Omega|}+b^2_z(\beta)t^2_1\beta^2_{31}\dfrac{q^2_z}{|\Omega|}}).
	\end{flalign}
	where we introduced the notation $q_p=\left(q^2_x+q^2_y\right)^{1/2}$, as well as the ratios of fermionic anisotropy parameters $\beta_{21}=t_2/t_1$,  $\beta_{31}=t_3/t_1$, and $\beta=\beta_{21}\beta_{31}$. The coefficients $\{a_p(\beta),a_z(\beta),b_p(\beta),b_z(\beta)\}$ are given in the Supplemental Material. Taking the static limit, the bosonic self-energy $\Pi(\boldsymbol{q},i\Omega\to0)$ depends linearly on momentum in the vicinity of the $\Gamma$ point of the Brillouin zone, thus it dominates the non-interacting bosonic propagator $D_3(\boldsymbol{q},i\Omega)=\left[c^2_p(q^2_x+q^2_y)+c^2_zq^2_z\right]^{-1}$. Hence,  the \textit{renormalized} bosonic propagator can be approximated as
	\begin{equation}\label{eq:RenormalizedBosonicPropagator}
		D_{3,r}(\boldsymbol{q},i\Omega)\approx\left[-\Pi(\boldsymbol{q},i\Omega)\right]^{-1}.
	\end{equation}
	Such an admission implies that the bosonic spectrum is strongly renormalized leading to a bosonic anomalous dimension $\eta^{(N_f\to\infty)}_\phi\approx0.5$ \cite{Abrikosov1974}. The one-loop frequency-dependent fermionic self-energy writes
	\begin{equation}\label{eq:LargeNfFermionicSelfEnergy}
		\dfrac{\Sigma(\bold{k},i\omega)}{e^2T}=\sum_{i\Omega}\int\dfrac{d^3q}{(2\pi)^3}\,G_3(\bold{k}+\bold{q},i\omega+i\Omega)D_{3,r}(\bold{q},i\Omega).
	\end{equation}
	In the zero-temperature limit, we introduce the chemical potential $\mu$ and parameter $\Lambda_0$ as the appropriate IR and UV cutoffs, respectively, to treat the emerging logarithmic divergence in the frequency integral of Eq.~\eqref{eq:LargeNfFermionicSelfEnergy}. The fermionic self-energy is then estimated to be
    \begin{flalign}\label{eq:LargeNfFermionicSelfEnergyResult}
    	&\Sigma(\bold{k},i\omega)=\dfrac{\text{ln}\,\Lambda_0/\mu}{N_f}\Big[-\Sigma_2\left(\beta_{21},\beta_{31}\right)t^2_2(k^2_x+k^2_y)\sigma_3\Big.\nonumber\\
    	&\Big.+\Sigma_3\left(\beta_{21},\beta_{31}\right)t^2_3k^2_z\sigma_3+\Sigma_\omega\left(\beta_{21},\beta_{31}\right)i\omega\mathbb{1}_2\nonumber\\
    	&\Big.-\Sigma_1\left(\beta_{21},\beta_{31}\right)2t^2_1k_z(k_x\sigma_1+k_y\sigma_2)\Big].
    \end{flalign}
    The calculation details are provided in Supplemental Material. Using Eq.~\eqref{eq:LargeNfFermionicSelfEnergyResult}, we construct the differential equations describing the evolution of the ratios $\{\beta^2_{21},\beta^2_{31}\}$ with respect to the dimensionless parameter $\ell=\Lambda_0/\mu$:
    \begin{equation}\label{eq:LargeNfFlowEquations}
    	\dfrac{d\text{ln}\beta^2_{21,31}}{d\text{ln}\ell}=\dfrac{\Sigma_{2,3}(\beta_{21},\beta_{31})-\Sigma_1(\beta_{21},\beta_{31})}{N_f}.
    \end{equation}
  	The expressions for the functions $\Sigma_{1,2,3}(\beta_{21},\beta_{31})$ are given in the Supplemental Material. In the latter, the numerical fixed-point solution corresponds to the strong anisotropy limit ${\beta^2_{21}}^*\ll1$ and ${\beta^2_{31}}^*\gg1$ implying that, after an infinite number of scale transformations, the fermionic anisotropy parameter $t^2_2$ is reduced to an infinitesimal value whereas $t^2_3$ becomes extremely large and $t^2_1$ remains non-zero but finite. Thus, the presence of long-range Coulomb interactions induces an emerging spatial anisotropy in the quadratic semimetallic phase. Furthermore, due to the non-zero renormalization of the fermionic Matsubara frequency in Eq.~\eqref{eq:LargeNfFermionicSelfEnergyResult}, the fermionic anomalous dimension at the strong anisotropy limit reads $\eta^{(N_f\to\infty)}_\psi=\Sigma_{\omega}(\beta_{21},\beta_{31})/(2N_f)\approx0.034/N_f$. Hence, the corresponding fermionic spectral density function does \textit{not} constitute of Dirac delta distributions exhibiting a sharp pole at particular energies, but rather of Lorentzian distributions around those energies. Such a renormalization pattern leads to the \textit{breakdown} of the Fermi-liquid behavior \cite{Lohneysen2007}.
  	\par\textit{Renormalization-group analysis}.~The conclusions from the large-$N_f$ analysis can be further corroborated and enriched through the application of the Wilsonian renormalization-group scheme \cite{Herbut2007}. We employ the latter by generalizing the spatially three-dimensional action in Eq.~\eqref{eq:ActionThreeDimensions} to a $(2+d)$-dimensional one by extending the one-dimensional rotational axis $\hat{z}$, about which the $C^z_4$ symmetry is enforced, to a $d$-dimensional manifold \cite{Han2019}. With the aim to extrapolate any findings to the physical case $d=1$, we introduce the real-valued dimension $\epsilon=2-d$ lying in the interval $[0,1]$. The resulting non-interacting fermionic and bosonic Hamiltonians read, respectively,
	\begin{align}\label{eq:QuadraticDHamiltonian}
		H_4(\bold{k})&= 2t^2_1k_z\qty[k_x\Gamma_1+k_y\Gamma_2]+2t^2_1k_4\qty[k_x\Gamma_3+k_y\Gamma_4]\nonumber\\
		&+\qty[t^2_2(k^2_x+k^2_y)-t^2_3(k^2_z+k^2_4)]\Gamma_5,
	\end{align}
	and
	\begin{equation}\label{eq:BosonicDHamiltonian}
		H_{4,b}(\bold{q})= c^2_p(q^2_x+q^2_y)+c^2_z(q^2_z+q^2_4).
	\end{equation}
	The renormalization-group scheme is applied along the radius of the extended $d$-dimensional manifold and the corresponding angular content is considered in two dimensions. Such a choice naturally generates the artificial fermionic (bosonic) momentum $k_4$ $(q_4)$. For the non-interacting Hamiltonian $H_4(\bold{k})$ to accommodate four-dimensional form factors, we have extended the representation of the Clifford algebra from two to four dimensions using the Dirac matrices $\{\mathbb{1}_4,\dots,\Gamma_{14}\}$ respecting the definitive relations $\{\Gamma_i,\Gamma_j\}=4\delta_{ij}\mathbb{1}_4$ and $\text{Tr}\qty[\Gamma_i\Gamma_j]=4\delta_{ij}$ with $i,j=\{1,\dots,14\}$ \cite{Snygg1997}. In terms of a characteristic length scale $L$, we introduce different scaling for the individual coordination axes $\hat{x}$ and $\hat{y}$ compared to the norm in $(x,y)$-plane, $[k_{x,y}]=L^{-f_{x,y}}$ and $[k^2_x+k^2_y]=L^{-f_{x+y}}$, due to the anisotropic scaling expected to occur among the fermionic parameters $\{t^2_1,t^2_2,t^2_3\}$ for strong long-range Coulomb interactions (from the large-$N_f$ analysis). Similarly, for the bosonic degrees of freedom, we consider the engineering dimension $[q^2_x+q^2_y]=L^{-b_{x+y}}$. Therefore, considering $z$ as the dynamic critical exponent \cite{Hohenberg1977}, the engineering dimensions of the renormalized quantities are
	 \begin{flalign}\label{eq:EngineerinDimensions}
	 	&[t^2_1]=L^{1+f_{x,y}-z},\qquad[t^2_2]=L^{f_{x+y}-z},\qquad[t^2_3]=L^{2-z},\nonumber\\
	 	&[\psi^2]=L^{f_{x+y}+d+2z},\qquad[c^2_p]=L^{b_{x+y}-z},\qquad[c^2_z]=L^{2-z},\nonumber\\
	 	&[\phi^2]=L^{b_{x+y}+d+2z},\qquad[e^2]=L^{d+b_{x+y}-2z}.
	\end{flalign}
	Since the non-interacting Hamiltonian $H_4(\bold{k})$ is the starting action for the microscopic fermionic theory, parameters $\{t^2_1,t^2_2,t^2_3\}$ are initially assumed \textit{marginal}. We thus set $f_{x+y}=b_{x+y}=2f_{x,y}=z=2$ leading the squared electronic charge $e^2$ to be a relevant perturbation at three spatial dimensions. Upon integrating out the high-energy modes lying in the radial interval $[\Lambda/\ell,\Lambda]$ in the $d$-dimensional manifold, with $\ell$ being the dimensionless scaling parameter and $\Lambda$ the momentum UV cutoff, the three-dimensional bosonic self-energy at one-loop order reads
	 \begin{flalign}\label{eq:RGBosonicSelfEnergy}
	 	\Pi(\bold{q})=&-N_f\,\alpha\,\text{ln}\ell\,B(\gamma_{21},\gamma_{31})\,c^2_p(q^2_x+q^2_y)\nonumber\\
	 	&-N_f\,\alpha\,\text{ln}\ell\,B(\gamma_{31},\gamma_{21})\,c^2_zq^2_z,
	 \end{flalign}
     where $B(\gamma_{21},\gamma_{31})=\gamma_{21}/2+\gamma^{-1}_{31}$, and we have introduced the dimensionless parameters
    \begin{equation}\label{eq:GammaParameters}
    	\gamma_{21}=\dfrac{c_z}{c_p}\beta^2_{21},\quad\gamma_{31}=\dfrac{c_p}{c_z}\beta^2_{31},\quad\alpha=\dfrac{e^2\Lambda^{d-4}}{24\pi t^2_1c_pc_z}.
    \end{equation}
    For the three-dimensional fermionic self-energy we find
    \begin{flalign}\label{eq:RGFermionicSelfEnergy}
    	\Sigma(\bold{k})=&-\alpha\,\text{ln}\ell\,F_1(\gamma_{21},\gamma_{31})\,2t^2_1k_z(k_x\sigma_1+k_y\sigma_2)\Big.\nonumber\\
    	&-\alpha\,\text{ln}\ell\,F_2(\gamma_{21},\gamma_{31})\,t^2_2(k^2_x+k^2_y)\sigma_3\nonumber\\
    	&+\alpha\,\text{ln}\ell\,F_3(\gamma_{21},\gamma_{31})\,t^2_3k^2_z\sigma_3,
    \end{flalign}
    with the functions $F_{1,2,3}(\gamma_{21},\gamma_{31})$ given in the Supplemental Material. Since the bosonic and fermionic self-energy expressions in Eqs.~\eqref{eq:RGBosonicSelfEnergy}--\eqref{eq:RGFermionicSelfEnergy} are controlled by the parameters $\{\alpha,\gamma_{21},\gamma_{31}\}$, we construct the corresponding renormalization-group (RG) equations
    \begin{flalign}\label{eq:RGEquations}
    	\dfrac{d\text{ln}\gamma_{21,31}}{d\text{ln}\ell}=&\pm N_f\alpha\dfrac{B(\gamma_{31},\gamma_{21})-B(\gamma_{21},\gamma_{31})}{2}\nonumber\\
    	&+\alpha\Big[F_{2,3}(\gamma_{21},\gamma_{31})-F_1(\gamma_{21},\gamma_{31})\Big],\nonumber\\
    	\dfrac{d\text{ln}\alpha}{d\text{ln}\ell}=&\epsilon-N_f\alpha\dfrac{B(\gamma_{31},\gamma_{21})+B(\gamma_{21},\gamma_{31})}{2}\nonumber\\
    	&-\alpha F_1(\gamma_{21},\gamma_{31}).
    \end{flalign}
    At the physical case $\epsilon=N_f=1$, the evolution of parameters $\{\alpha,\gamma_{21},\gamma_{31}\}$ in the limit of infinite scale transformations leads to the fixed-point values (see Supplemental Material)
    \begin{equation}\label{eq:StableInteractingFixedPoint}
    	(\alpha^*,\gamma^*_{21},\gamma^*_{31})|^{N_f=1}_{\epsilon=1}\approx(0.549,1.129,1.894).
    \end{equation}
   That result implies the existence of a stable interacting fixed point. The fixed-point value for the coupling constant $\alpha^*|^{N_f=1}_{\epsilon=1}\propto\epsilon/2$ is in agreement with the expectation that the renormalized coupling constant at one-loop order is proportional to the distance from the \textit{upper critical dimension} $(\epsilon=0)$, and that the fixed-point value is dominated by the renormalization of the bosonic spectrum \cite{Abrikosov1974}. In Fig.~\ref{fig:PhasePortrait}, the stable character of the interacting fixed point is confirmed by the phase portrait of parameters $\{\gamma_{21},\gamma_{31}\}$ for the value $\alpha^*=0.549$. Regarding the RG equations for the ratios $\{\beta^2_{21},\beta^2_{31}\}$ at the interacting fixed point, we find
	\begin{equation}
		\left(\dfrac{1}{\beta^2_{21}}\dfrac{d\beta^2_{21}}{d\text{ln}\ell},\dfrac{1}{\beta^2_{31}}\dfrac{d\beta^2_{31}}{d\text{ln}\ell}\right)|^{N_f=1}_{\epsilon=1}\approx(-0.203,0.203),
	\end{equation}
	implying that parameter $\beta^2_{21}$ $(\beta^2_{31})$ is negatively (positively) diverging as a function of the scale transformations. Thus, $\beta^2_{21}$ becomes \textit{irrelevant} parameter at the stable interacting fixed point, and the existence of the latter does \textit{not} depend on the presence of the fermionic anisotropy parameter $t^2_2$, and in extent, on the term $(k^2_x+k^2_y)\sigma_z$ in the fermionic Hamiltonian. Such a conclusion consolidates the argument made within the large-$N_f$ analysis (see Eq.~\eqref{eq:LargeNfFlowEquations}) about the emergence of spatial anisotropy in the presence of Coulomb interactions. At the interacting fixed point, the one-loop renormalized dynamic critical exponent $\eta_\omega$ is
	\begin{flalign}\label{eq:RenormalizedDynamicCriticalExponent}
		&\eta_\omega|^{N_f=1}_{\epsilon=1}=z-\alpha^*F_1(\gamma^*_{21},\gamma^*_{31})|^{N_f=1}_{\epsilon=1}\approx1.803
	\end{flalign}
	resulting to an anisotropic scaling among momentum and energy degrees of freedom i.e.~if momentum has engineering dimensions $L^{-2}$ for a quadratic semimetal, then the energy has engineering dimensions $L^{-\eta_\omega}$. The one-loop fermionic anomalous dimension $\eta_\psi$ is estimated to be
	\begin{flalign}\label{eq:RenormalizedFermionicAnomalousDimension}
		\eta_\psi|^{N_f=1}_{\epsilon=1}=\alpha^*F_1(\gamma^*_{21},\gamma^*_{31})|^{N_f=1}_{\epsilon=1}\approx0.197
	\end{flalign}
	implying that the fermionic Green's function obtains a non-zero imaginary part. Finally, the one-loop bosonic anomalous dimension $\eta_\phi$ reads
	\begin{flalign}\label{eq:RenormalizedBosonicAnomalousDimension}
		\eta_\phi|^{N_f=1}_{\epsilon=1}=&\dfrac{\alpha^*}{4}\Big[2F_1(\gamma^*_{21},\gamma^*_{31})+N_fB(\gamma^*_{21},\gamma^*_{31})\Big.\nonumber\\
		&\Big.+N_fB(\gamma^*_{31},\gamma^*_{21})\Big]|^{N_f=1}_{\epsilon=1}\approx 0.5.
	\end{flalign}
	 Therefore, the renormalized bosonic Hamiltonian is expected to scale as $L^{-1}$ indicating a strong renormalization of the bosonic spectrum. Qualitatively, the results in Eqs.~\eqref{eq:RenormalizedFermionicAnomalousDimension}--\eqref{eq:RenormalizedBosonicAnomalousDimension} agree with the conclusions derived using the large-$N_f$ method, thus reinforcing the argument for the breakdown of the Fermi-liquid behavior at the stable interacting fixed point.
	\par\textit{Physical observables}.~A Berry-dipole semimetal has been recently realized in a three-dimensional phononic lattice \cite{Mo2025}, in which long-range Coulomb interactions can be tuned upon considering polar acoustic resonators \cite{Li2024}.~Furthermore, the high tunability of nitrogen-vacancy centers as well as topoelectric circuits might also serve as candidate platforms for realizing Coulomb interactions in a Berry-dipole semimetal \cite{Zhou2020,Sahin2025}. It is then natural to further characterize the interacting semimetallic phase through the dependence of the density of states $\rho(\nu)$, specific heat $C_v(T)$, compressibility $\kappa(T)$, diamagnetic susceptibility $\chi(T)$, static and high-temperature optical conductivity, $\sigma^\text{DC}(T)$ and $\sigma^{T\to\infty}(\nu)$, on the probed energy $\nu$ or temperature scale $T$. We first determine the scaling form of those observables with respect to the energy or temperature and the three fermionic anisotropy parameters $\{t^2_1,t^2_2,t^2_3\}$  (see Supplemental Material):
	\begin{flalign}\label{eq:PhysicalObservables}
		&\rho(\nu)\propto\dfrac{|\nu|^{1/2}}{t^2_2t_3},\qquad C_v(T)\propto\dfrac{T^{3/2}}{t^2_2t_3},\qquad\kappa(T)\propto\dfrac{T^{1/2}}{t^2_2t_3},\nonumber\\
		&\chi_{x,\,y}(T)\propto T^{1/2}t_3,\qquad \chi_z(T)\propto T^{1/2}\dfrac{t^4_1}{t^3_3},\nonumber\\
		&\sigma^\text{DC}_{xx,\,yy}(T)\propto \dfrac{T^{3/2}}{t_3},\qquad \sigma^\text{DC}_{zz}(T)\propto T^{3/2}\dfrac{t^3_3}{t^4_1},\nonumber\\
		&\sigma^{T\to\infty}_{xx,yy}(\nu)\propto \dfrac{\omega^{1/2}}{t_3},\qquad \sigma^{T\to\infty}_{zz}(\nu)\propto\nu^{1/2}\dfrac{t^3_3}{t^4_1}.
	\end{flalign}
	The emerging spatial anisotropy at the interacting fixed point is expected to induce power-law corrections to the corresponding observables \cite{Sheehy2007,Isobe2016,Han2019}. Therefore, we consider the RG equations for the fermionic anisotropy parameters \textit{at} the interacting fixed point:
	     \begin{figure}[t]
		\includegraphics[scale=0.7]{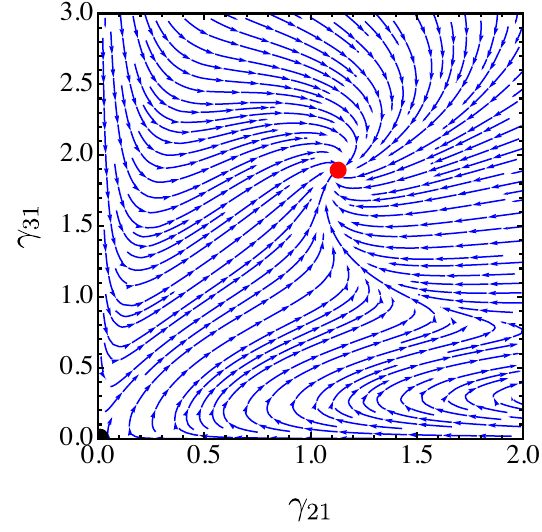}
		\caption{Phase portrait at $\epsilon=N_f=1$ of parameters $\{\gamma_{21},\gamma_{31}\}$ defined in Eq.~\eqref{eq:GammaParameters} for the fixed-point value $\alpha^*=0.549$. The red dot denotes the stable interacting fixed point and the black dot is the Gaussian non-interacting fixed point.}
		\label{fig:PhasePortrait}
	\end{figure}
	\begin{flalign}\label{eq:RGEquationsFermionicAnisotropyParametersInteractingFixedPoint}
		&\dfrac{1}{t^2_1}\dfrac{dt^2_1}{d\text{ln}\ell}|^{N_f=1}_{\epsilon=1}=z^*-f^*_{x,y}-1+\alpha^*F_1(\gamma^*_{21},\gamma^*_{31})|^{N_f=1}_{\epsilon=1}\stackrel{!}{=}0,\nonumber\\
		&\dfrac{1}{t^2_2}\dfrac{dt^2_2}{d\text{ln}\ell}|^{N_f=1}_{\epsilon=1}=z^*-f^*_{x+y}+\alpha^*F_2(\gamma^*_{21},\gamma^*_{31})|^{N_f=1}_{\epsilon=1}\stackrel{!}{=}0,\nonumber\\
		&\dfrac{1}{t^2_3}\dfrac{dt^2_3}{d\text{ln}\ell}|^{N_f=1}_{\epsilon=1}=z^*-\alpha^*F_3(\gamma^*_{21},\gamma^*_{31})|^{N_f=1}_{\epsilon=1}\stackrel{!}{=}0.
	\end{flalign}
	Parameters $\{t^2_1,t^2_2,t^2_3\}$ are no longer marginal and they rather exhibit the engineering dimensions given in Eq.~\eqref{eq:EngineerinDimensions} with exponents determined via Eq.~\eqref{eq:RGEquationsFermionicAnisotropyParametersInteractingFixedPoint}. Consequently, the observables in Eq.~\eqref{eq:PhysicalObservables} are modified as
	\begin{flalign}\label{eq:PhysicalObservablesInteractingFixedPoint}
		&\rho(\nu)|^{N_f=1}_{\epsilon=1}\propto|\nu|^{1/2+\eta_1},\qquad C_v(T)|^{N_f=1}_{\epsilon=1}\propto T^{3/2+\eta_1},\nonumber\\
		&\kappa(T)|^{N_f=1}_{\epsilon=1}\propto T^{1/2+\eta_1},\qquad \chi_{x,y}(T)|^{N_f=1}_{\epsilon=1}\propto T^{1/2-\eta_2},\nonumber\\
		&\chi_z(T)|^{N_f=1}_{\epsilon=1}\propto T^{1/2+\eta_3},\qquad\sigma^{\text{DC}}_{xx,yy}(T)|^{N_f=1}_{\epsilon=1}\propto T^{3/2+\eta_2},\nonumber\\
		&\sigma^{\text{DC}}_{zz}(T)|^{N_f=1}_{\epsilon=1}\propto T^{3/2-\eta_3},\qquad\sigma^{T\to\infty}_{xx,yy}(\nu)|^{N_f=1}_{\epsilon=1}\propto\nu^{1/2+\eta_2},\nonumber\\
		&\sigma^{T\to\infty}_{zz}(\nu)|^{N_f=1}_{\epsilon=1}\propto\nu^{1/2-\eta_3},
	\end{flalign}
	where $\left(\eta_1,\eta_2,\eta_3\right)=\left(0.121,0.125,0.129\right)$. Further computation details are provided in the Supplemental Material. The interaction-induced spatial anisotropy can thus be signaled by the non-trivial temperature and energy scaling of the ratios of diamagnetic susceptibilities $\chi_z(T)/\chi_{x,\,y}(T)|^{N_f=1}_{\epsilon=1}\propto T^{\eta_3+\eta_2}$ and optical conductivities $\sigma^{a}_{xx,\,yy}(b)/\sigma^{a}_{zz}(b)|^{N_f=1}_{\epsilon=1}\propto b^{\eta_2+\eta_3}$, where $a=\{\text{DC},\,T\to\infty\}$ and $b=\{T,\,\nu\}$. Overall, the modified energy and temperature scaling highlights the deviation of the behavior of the considered obervables from the estimations predicted within the context of Fermi-liquid theory, hence they indicate that the emergent spatial anisotropy at the interacting fixed point is accompanied by a \textit{non-Fermi-liquid behavior} \cite{Lohneysen2007}.
	\par\textit{Fate of flux of Berry curvature}.~For the spatially three-dimensional Hamiltonian $H_3(\bold{k})$ given in Eq.~\eqref{eq:QuadraticHamiltonian}, the fluxes of Berry curvature $C_{z<0,\,z>0}$ through the lower- and upper-half $z$-planes, respectively, evolve as functions of the dimensionless scaling parameter $\ell$ according to the RG equations (see Supplemental Material)
	\begin{equation}\label{eq:RGEquationFluxBerryCurvature}
		\dfrac{dC_{z<0,\,z>0}}{d\text{ln}\ell}=\pm\dfrac{1}{2}\Big(\dfrac{d\beta^2_{21}}{d\text{ln}\ell}+\dfrac{d\beta^2_{31}}{d\text{ln}\ell}\Big).
	\end{equation}
	In the non-interacting limit, the fermionic anisotropy parameters are marginal, therefore Eq.~\eqref{eq:RGEquationFluxBerryCurvature} implies that the fluxes $C_{z<0,\,z>0}$ are independent of scale transformations, and the admitted Berry-dipole semimetal shows a quantized flux through the corresponding half $z$-planes \cite{Zhuang2024}. However, in the presence of long-range Coulomb interactions, the fermionic anisotropy parameters become scale-dependent, and the evolution of the fluxes $C_{z<0,\,z>0}$ at the interacting fixed point indicates a divergent behavior:
	\begin{equation}
		C_{z<0,\,z>0}|^{N_f=1}_{\epsilon=1}\to\pm\infty.
	\end{equation}
	 The Berry curvature is therefore greatly enhanced and the resulting interacting fixed point can be termed \textit{topological}, though the quantization of the flux through the half $z$-planes is destroyed. We note that the fluxes $C_{z<0,\,z>0}$ are linearly related to both the intrinsic and extrinsic non-linear Hall conductivities, i.e.~$\sigma^\text{int}_{xxz},\,\sigma^\text{ext}_{zxx}\sim|C_{z<0,\,z>0}|$ \cite{Zhuang2023}. Hence, the fate of the flux of Berry curvature is manifested in the non-linear Hall response leading to the following observational criterion: as a function of the long-range Coulomb interaction, the Berry-dipole semimetal undergoes a quantum phase transition into a topological anisotropic non-Fermi liquid when both the intrinsic and extrinsic non-linear Hall conductivities show giant enhancement.
	 \par\textit{Conclusion}.~In summary, we proposed the existence of a novel semimetallic phase of matter, lying beyond the predictive power of known classification rules, that features spatially anisotropic non-Fermi liquid behavior with an enhanced topology in terms of large fluxes of Berry curvature in the upper- and lower-half $z$-planes. The non-Fermi liquid behavior can be experimentally probed in acoustic polar lattices and, potentially, in nitrogen-vacancy centers and topoelectric circuits, by measuring the power-law corrections to the density of states, specific heat, compressibility, diamagnetic and optical conductivities. The generated topological features can be accessed through measurements of the intrinsic and extrinsic non-linear Hall conductivities.
	\par\textit{Acknowledgements}.~K.L.~is grateful to S.~Sur for discussions in the early parts of the work, and to I.F.~Herbut as well as M.M.~Scherer for carefully reading and providing helpful comments on earlier versions of this manuscript. This work is financially supported by the Deutsche Forschungsgemeinschaft (DFG) through the Walter Benjamin program (LA5962/2-1, Project id No.~567023139).

\renewcommand{\emph}{\textit}
\bibliography{NFLBerryDipole}

\end{document}


\title{Supplemental Material: Topological anisotropic non-Fermi liquid from a Berry-dipole semimetal}

\author{Konstantinos Ladovrechis}
\affiliation{ 
		Theoretische Physik III, Ruhr-Universität Bochum, D-44801 Bochum, Germany}	
		
\maketitle		


\section{Details of the large-$N_f$ calculation}
In the present section of the Supplemental Material, we provide the computation details involved in the employment of the large-$N_f$ method in the corresponding section of the main text. We consider the following three-dimensional fermionic Hamiltonian in momentum domain lying at the topological quantum critical point (TQCP) separating the Hopf insulator from a trivial insulator \cite{Moore2008,Zhuang2024}
\begin{align}\label{eq:QuadraticHamiltonian}
	H_3(\bold{k}) =\sum^3_{i=1}d_i(\bold{k})\sigma_i,
\end{align}
where $\{\sigma_1,\sigma_2,\sigma_3\}$ are the three Pauli matrices and the form factors read $d_1(\bold{k})=2t^2_1k_xk_z$, $d_2(\bold{k})=2t^2_1k_yk_z$ and $d_3(\bold{k})=t^2_2(k^2_x+k^2_y)-t^2_3k^2_z$. The spectrum is $E_3(\bold{k})=\sqrt{\sum^3_{i=1}d_i(\bold{k})}$. The non-interacting fermionic Matsubara Green's function writes
\begin{equation}\label{eq:FermionicMatsubaraGreen'sFunction}
	G_3(\bold{k},i\omega)=\left[-i\omega\mathbb{1}_2+H_3(\bold{k})\right]^{-1}.
\end{equation}
Regarding the real bosonic field $\phi(\bold{q},i\Omega)$, the corresponding Hamiltonian assumes the form
\begin{align}\label{eq:BosonicHamiltonian}
	H_b(\bold{q}) =c^2_p(q^2_x+q^2_y)+c^2_zq^2_z,
\end{align}
and the non-interacting bosonic Matsubara Green's function is
\begin{equation}
	D_3(\bold{q},i\Omega)=\left[-i\Omega+H_b(\bold{q})\right]^{-1}.
\end{equation}
\subsection{Bosonic self-energy}
Considering the fermionic flavour number $N_f$, the one-loop bosonic self-energy is given by the expression \cite{FetterWalecka}
\begin{flalign}\label{eq:LargeNfBosonicSelfEnergy}
	\Pi(\boldsymbol{q},i\Omega)&=N_fe^2T\sum_{i\omega}\int\dfrac{d^3k}{(2\pi)^3}\Tr\left[G_3(\bold{k}+\bold{q},i\omega+i\Omega)G_3(\bold{k},i\omega)\right]\nonumber\\
	&=-N_fe^2\int\dfrac{d^3k}{(2\pi)^3}\dfrac{E_3(\bold{k}+\bold{q})+E_3(\bold{k})}{\Omega^2+(E_3(\bold{k}+\bold{q})+E_3(\bold{k}))^2}\left[1-\dfrac{\sum^3_{i=1}d_i(\bold{k}+\bold{q})d_i(\bold{k})}{E_3(\bold{k}+\bold{q})E_3(\bold{k})}\right].
\end{flalign}
We will proceed by constructing an ansatz for the bosonic self-energy. To that end, we first study the form of the latter for the bosonic momenta $q_{x,y}$ lying perpendicular to the rotational $\hat{z}$ axis.
\subsubsection{$q_{x,y}$-dependence}\label{sec:LargeNfBosonicSelfEnergyQp}
\begin{figure}[t]
	\centering 	\captionsetup{justification=centering}\includegraphics[width=0.35\textwidth]{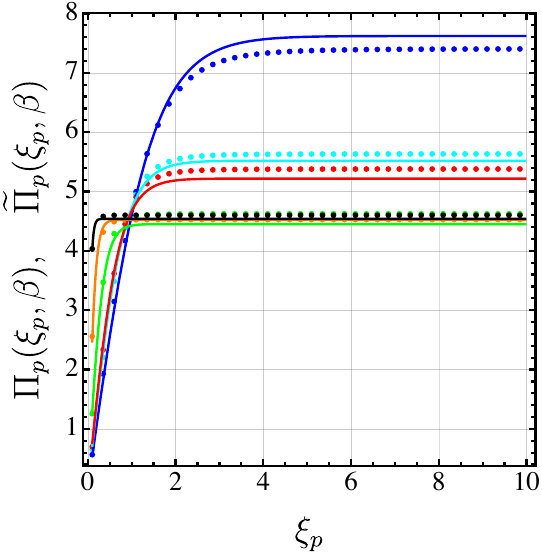}
	\caption{The expression $\Pi_p(\xi_p,\beta)\equiv\Pi_p(q_p,\xi_p,\phi=\pi/3,\beta,t_1)/\left(-(N_fe^2q_p)/((2\pi)^3t^2_1)\right)$ in Eq.~\eqref{eq:BosonicSelfEnergyQp} (dotted data points) is plotted against the ansatz $\widetilde{\Pi}_p(\xi_p,\beta)\equiv\widetilde{\Pi}_p(q_p,\xi_p,\beta,t_1)/\left(-(N_fe^2q_p)/((2\pi)^3t^2_1)\right)$ given in Eq.~\eqref{eq:AnsatzLargeNfBosonicSelfEnergyQp} (continuous lines) as functions of the amplitude $\xi_p\in[0.1,10]$ for different values of the parameter $\beta$: $\beta=0.5$ (blue), $\beta=1$ (cyan), $\beta=1.156$ (red), $\beta=2.5$ (green), $\beta=5$ (orange) and $\beta=10$ (black). The value $\beta=1.156$ has been chosen because it is connected to the interacting fixed point derived from the $\epsilon$-expansion within the renormalization-group scheme (see main text): $\beta|^{N_f=1}_{\epsilon=1}=\sqrt{\gamma^*_{21}\gamma^*_{31}}|^{N_f=1}_{\epsilon=1}\approx1.156$.}
	\label{fig:ComparingAnsatzNumericalIntegrationLargeNfBosonicSelfEnergyQp}
\end{figure}

Upon making the variable transformations $k_x=xq_p\cos\phi$, $k_y=yq_p\cos\phi$ and $k_z=zq_p/\beta^2_{31}$ with $\phi=\arctan\left(q_y/q_x\right)$, $q_p=\sqrt{q^2_x+q^2_z}$, and $\beta_{21,31}=t_{2,3}/t_1$ being ratios of fermionic anisotropy parameters, the bosonic self-energy expression reads
\begin{flalign}\label{eq:BosonicSelfEnergyQp}
	\Pi_p(q_p,\xi_p,\phi,\beta,t_1)=&-\dfrac{N_fe^2q_p}{(2\pi)^3t^2_1}\int^\infty_{-\infty}dx\int^\infty_{-\infty}dy\int^\infty_{-\infty}dz\dfrac{\sqrt{\pi_1(x,y,z,\beta)}+\sqrt{\pi_2(x,y,z,\phi,\beta)}}{\left(\sqrt{\pi_1(x,y,z,\beta)}+\sqrt{\pi_2(x,y,z,\phi,\beta)}\right)^2+\left(\dfrac{1}{\xi_p}\right)^2}\times\nonumber\\
	&\times\left[1-\dfrac{x^2 \beta^4 + x^4 \beta^4 + y^2\beta^4 + 2 x^2 y^2 \beta^4 + y^4 \beta^4 + 2 x (x^2 \beta^4 + y^2 \beta^4)\cos \phi + 2 y (x^2 \beta^4 + y^2 \beta^4)\sin\phi}{\sqrt{\pi_1(x,y,z,\beta)}\sqrt{\pi_2(x,y,z,\phi,\beta)}}\right],
\end{flalign}
where we have introduced the parameters $\xi_p=t^2_1q^2_p/\left(\beta^2_{31}\Omega\right)$ and  $\beta=\beta_{21}\beta_{31}$ as well as the functions
\begin{flalign}
	&\pi_1(x,y,z,\beta)=(x^2 + y^2)^2 \beta^4,\nonumber\\
	&\pi_2(x,y,z,\phi,\beta)=\beta^4 (1 + x^2 + y^2 + 2 x \cos\phi + 2 y \sin\phi)^2.
\end{flalign}
The integration form of Eq.~\eqref{eq:BosonicSelfEnergyQp} can be approximated by the ansatz
\begin{flalign}\label{eq:AnsatzLargeNfBosonicSelfEnergyQp}
	&\widetilde{\Pi}_p(q_p,\xi_p,\beta,t_1)=-\dfrac{N_fe^2q_p}{(2\pi)^3t^2_1}a_p(\beta)\tanh(b_p(\beta)|\xi_p|)
\end{flalign}
where the curve-fitting parameters $\{a_p(\beta),b_p(\beta)\}$ are found to be
\begin{flalign}
	&a_p(\beta)=30.5397\left[1+0.221845\dfrac{\beta^{1/2}+\beta^{1/3}+\beta^{1/4}+\beta^{1/5}}{0.303172-\text{ln}(3+\beta)}\right],\nonumber\\
	&b_p(\beta)=0.02898\left[1+34.8242\dfrac{\beta^{2/3}+\beta^{1/2}-0.891338\beta^{1/3}}{1-0.106504+0.00609918\beta^2-0.000109586\beta^3}\right].
\end{flalign}
In Fig.~\ref{fig:ComparingAnsatzNumericalIntegrationLargeNfBosonicSelfEnergyQp}, the numerically integrated expression in Eq.~\eqref{eq:BosonicSelfEnergyQp} for $\phi=\pi/3$ is plotted against the ansatz given in Eq.~\eqref{eq:AnsatzLargeNfBosonicSelfEnergyQz} in terms of the parameters $\{\xi_p,\beta\}$. Qualitatively, the agreement between numerical data and the continuous function is satisfactory for the plotted $\beta$ values as well as throughout the $\xi_p$ interval. We also note that the resulting numerical integration does not depend on the azimuthial angle $\phi$, due to the presence of the $C^z_{4}$ symmetry about the $\hat{z}$ axis in the Hamiltonian $H_3(\bold{k})$. Next, we study the form of Eq.~\eqref{eq:LargeNfBosonicSelfEnergy} in the presence of bosonic momentum $q_z$ along the rotational $\hat{z}$ axis.
\subsubsection{$q^2_z$ dependence}
\begin{figure}[t]
	\centering\includegraphics[width=0.36\textwidth]{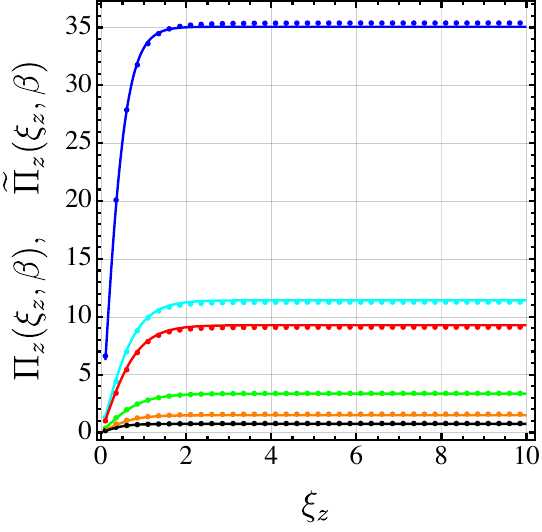}
	\caption{The expression $\Pi_z(\xi_z,\beta)\equiv\Pi_z(q_z,\xi_z,\beta,\beta_{31},t_1)//\left(-(N_fe^2|q_z|\beta^2_{31})/((2\pi)^3t^2_1)\right)$ in Eq.~\eqref{eq:BosonicSelfEnergyQz} (dotted data points) is plotted against the ansatz $\widetilde{\Pi}_z(\xi_z,\beta)\equiv\widetilde{\Pi}_z(q_z,\xi_z,\beta,\beta_{31},t_1)/\left(-(N_fe^2|q_z|\beta^2_{31})/((2\pi)^3t^2_1)\right)$ (continuous lines) as functions of the parameter $\xi_z\in[0.1,10]$ for different values of the parameter $\beta$: $\beta=0.5$ (blue), $\beta=1$ (cyan), $\beta=1.156$ (red), $\beta=2.5$ (green), $\beta=5$ (orange) and $\beta=10$ (black). The value $\beta=1.156$ has been chosen because it is connected to the interacting fixed point derived from the $\epsilon$-expansion within the renormalization-group scheme (see main text): $\beta|^{N_f=1}_{\epsilon=1}=\sqrt{\gamma^*_{21}\gamma^*_{31}}|^{N_f=1}_{\epsilon=1}\approx1.156$.}
	\label{fig:ComparingAnsatzNumericalIntegrationLargeNfBosonicSelfEnergyQz}
\end{figure}

Performing the variable transformations $k_x=xq_z\beta^2_{31}$, $k_y=yq_z\beta^2_{31}$ and $k_z=zq_z$, the corresponding bosonic self-energy expression reads
\begin{flalign}\label{eq:BosonicSelfEnergyQz}
	\Pi_z(q_z,\xi_z,\beta,\beta_{31},t_1)&=-\dfrac{N_fe^2|q_z|}{(2\pi)^3t^2_1}\beta^2_{31}\int^\infty_{-\infty}dx\int^\infty_{-\infty}dy\int^\infty_{-\infty}dz\dfrac{\sqrt{\pi_3(x,y,z,\beta)}+\sqrt{\pi_4(x,y,z,\beta)}}{\left(\sqrt{\pi_3(x,y,z,\beta)}+\sqrt{\pi_4(x,y,z,\beta)}\right)^2+\left(\dfrac{1}{\xi_z}\right)^2}\times\nonumber\\
	&\times\left[1+\dfrac{-z (1+z) (4 (x^2+y^2)+z+z^2)+(x^2+y^2) (1+2 z (1+z))\beta^2-(x^2+y^2)^2 \beta^4}{\sqrt{\pi_3(x,y,z,\beta)}\sqrt{\pi_4(x,y,z,\beta)}}\right],			
\end{flalign}
where we have introduced the parameter $\xi_z=t^2_3q^2_z/\Omega$ and the functions
\begin{flalign}
	&\pi_3(x,y,z,\beta)=4 (x^2 + y^2) z^2 + z^4 - 
	2 (x^2 + y^2) z^2 \beta^2 + (x^2 + y^2)^2\beta^4,\nonumber\\
	&\pi_4(x,y,z,\beta)=4 x^2 (1+z)^2+4 y^2 (1+z)^2+((1+z)^2-(x^2+y^2) \beta^2)^2.
\end{flalign}
The integration form of Eq.~\eqref{eq:BosonicSelfEnergyQz} can be approximated by the ansatz
\begin{flalign}\label{eq:AnsatzLargeNfBosonicSelfEnergyQz}
	&\widetilde{\Pi}_z(q_z,\xi_z,\beta,\beta_{31},t_1)=-\dfrac{N_fe^2|q_z|}{(2\pi)^3t^2_1}\beta^2_{31}a_z(\beta)\tanh(b_z(\beta)|\xi_z|)
\end{flalign}
where the curve-fitting parameters $\{a_z(\beta),b_z(\beta)\}$ are found to be
\begin{flalign}
	&a_z(\beta)=\dfrac{7.78245+10.6063\beta^2+3.55509\beta^3+0.900298\beta^{7/2}}{\beta^2+\beta^4},\nonumber\\
	&b_z(\beta)=\dfrac{0.0254895+0.112266\beta+0.197267\beta^2+0.0980718\beta^3-0.00187081\beta^4}{-0.96582\beta+0.627269\beta^2+\text{ln}(1+\beta)}.
\end{flalign}
In Fig.~\ref{fig:ComparingAnsatzNumericalIntegrationLargeNfBosonicSelfEnergyQz}, the numerically integrated expression in Eq.~\eqref{eq:BosonicSelfEnergyQz} is plotted against the ansatz given in Eq.~\eqref{eq:AnsatzLargeNfBosonicSelfEnergyQz} in terms of the parameters $\{\xi_z,\beta\}$. Similarly to the ansatz for the $q_{x,y}$-dependent bosonic self-energy, the agreement between numerical data and the continuous function is qualitatively satisfactory for the plotted values of the parameter $\beta$ as well as throughout the $\xi_z$ interval.
\subsubsection{Generic $q$ dependence}
\begin{figure}[t]
	\centering
	\subfloat[$\beta=0.5$]{\includegraphics[width=0.33\textwidth]{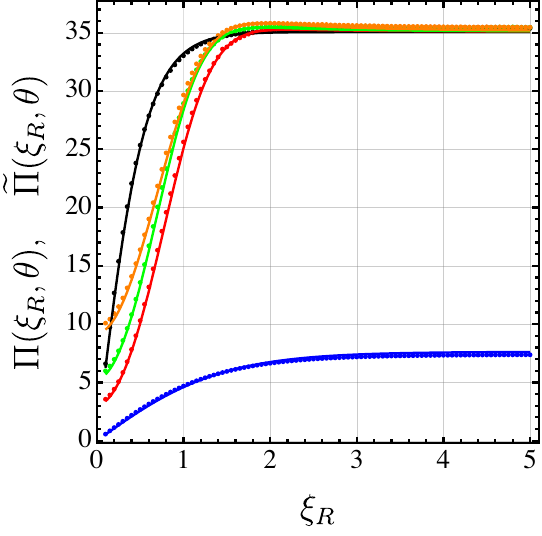}}
	\subfloat[$\beta=1$]{\includegraphics[width=0.33\textwidth]{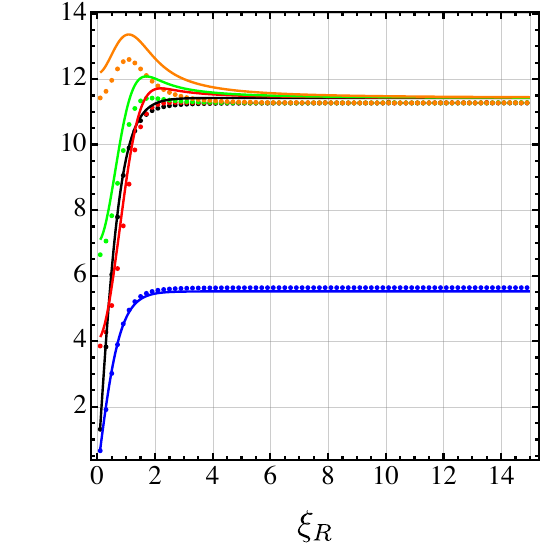}}
	\subfloat[$\beta=1.156$]{\includegraphics[width=0.33\textwidth]{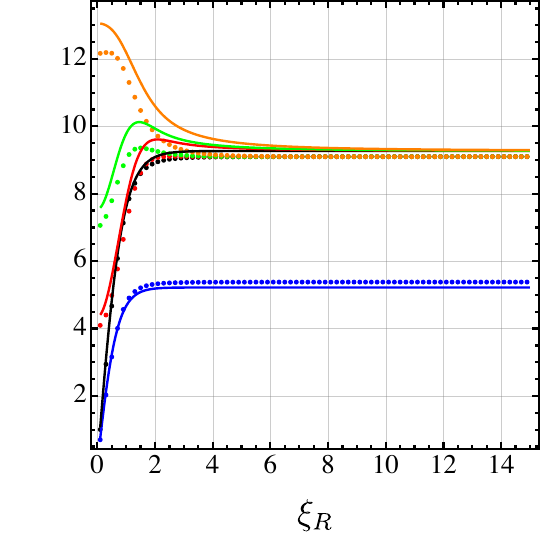}}\\
	\subfloat[$\beta=2.5$]{\includegraphics[width=0.33\textwidth]{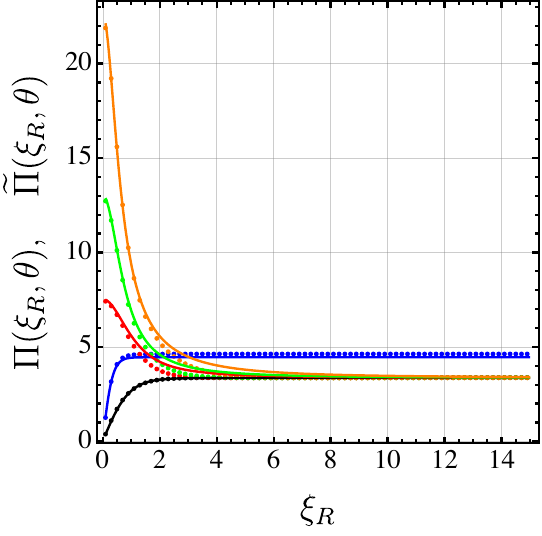}}
	\subfloat[$\beta=5$]{\includegraphics[width=0.33\textwidth]{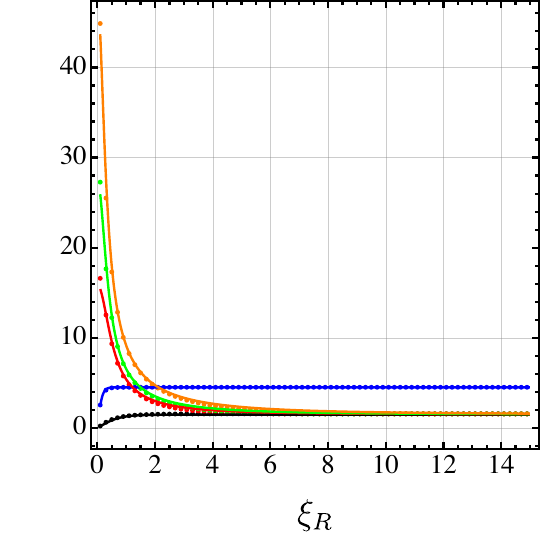}}
	\subfloat[$\beta=10$]{\includegraphics[width=0.33\textwidth]{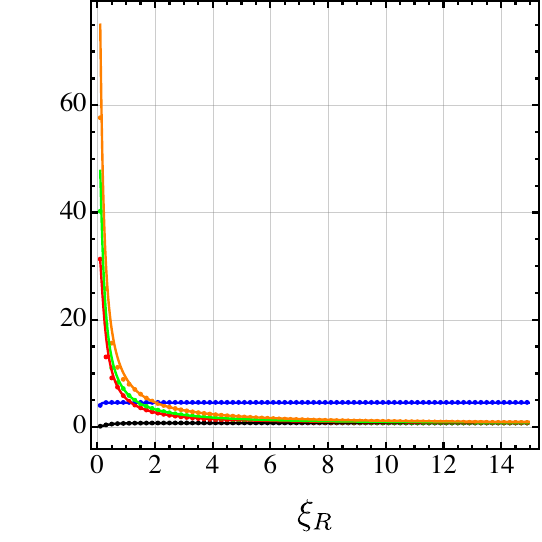}}
	\caption{The expression $\Pi(\xi_R,\theta,\beta)\equiv\Pi(q_z,\xi_R,\theta,\phi=\pi/3,\beta,\beta_{31},t_1)/\left(-(N_fe^2|q_z|\beta^2_{31})/((2\pi)^3t^2_1)\right)$ in Eq.~\eqref{eq::LargeNfBosonicSelfEnergyRewrittenXiParameters} (dotted data points) is plotted against the ansatz $\widetilde{\Pi}(\xi_R,\theta,\beta)\equiv\widetilde{\Pi}(q_z,\xi_R,\theta,\beta,\beta_{31},t_1)/\left(-(N_fe^2|q_z|\beta^2_{31})/((2\pi)^3t^2_1)\right)$ given in Eq.~\eqref{eq:AnsatzLargeNfBosonicSelfEnergyXiParameters} (continuous line) as functions of the amplitude $\xi_R\geq0.1$ and of different values of the parameter $\beta$: (a) $\beta=0.5$, (b) $\beta=1$, (c) $\beta=1.156$, (d) $\beta=2.5$, (e) $\beta=5$ and (f) $\beta=10$. The value $\beta=1.156$ has been chosen because it is connected to the interacting fixed point derived from the $\epsilon$-expansion within the renormalization-group scheme (see main text): $\beta|^{N_f=1}_{\epsilon=1}=\sqrt{\gamma^*_{21}\gamma^*_{31}}|^{N_f=1}_{\epsilon=1}\approx1.156$. At each plot, the color code corresponds to the following values for the angle $\theta$: $\theta=0$ (blue), $\theta=\pi/2$ (black), $\theta=\pi/3$ (red), $\theta=\pi/4$ (green) and $\theta=\pi/6$ (orange).}
	\label{fig:AnsatzLargeNfBosonicSelfEnergyBeta}
\end{figure}

Considering the case for the bosonic self-energy in Eq.~\eqref{eq:LargeNfBosonicSelfEnergy} with generic $q$-dependence, we make the variable transformations $k_x=xq_p\cos\phi$ and $k_y=yq_p\sin\phi$ with $\phi=\arctan\left(q_y/q_x\right)$ and $q_p=\sqrt{q^2_x+q^2_y}$ as well as $k_z=zq_z$ to obtain the expression
\begin{flalign}\label{eq::LargeNfBosonicSelfEnergyRewritten}
	\Pi(q_p,q_z,i\Omega,\phi,\beta,\beta_{31},t_1)&=-\dfrac{N_fe^2|q_z|}{(2\pi)^3t^2_1}\beta^2_{31}\int^\infty_{-\infty}dx\int^\infty_{-\infty}dy\int^\infty_{-\infty}dz\left[1+\dfrac{\pi_5\left(x,y,z,\phi,\beta,\beta^2_{31}\dfrac{q_z}{q_p}\right)}{\sqrt{\pi_6\left(x,y,z,\phi,\beta\right)}\sqrt{\pi_7\left(x,y,z,\phi,\beta\right)}}\right]\times\nonumber\\
	&\times\dfrac{\sqrt{\pi_6\left(x,y,z,\phi,\beta\right)}+\sqrt{\pi_7\left(x,y,z,\phi,\beta\right)}}{\left(\sqrt{\pi_6(\left(x,y,z,\phi,\beta\right)}+\sqrt{\pi_7\left(x,y,z,\phi,\beta\right)}\right)^2+\left(\dfrac{\beta^2_{31}\Omega}{t^2_1q^2_p}\right)^2},\nonumber\\
\end{flalign}
where we have introduced the functions
\begin{flalign}
	&\pi_5\left(x,y,z,\phi,\beta,\beta^2_{31}\dfrac{q_z}{q_p}\right)=x^2 \beta^4 + x^4 \beta^4 + y^2 \beta^4 + 2 x^2 y^2 \beta^4 + y^4 \beta^4 + 4 x^2 z\left(\beta^2_{31}\dfrac{q_z}{q_p}\right)^2 + 4 y^2 z\left(\beta^2_{31}\dfrac{q_z}{q_p}\right)^2 \nonumber\\
	&\qquad\qquad\qquad\qquad\qquad+ 4 y^2 z^2\left(\beta^2_{31}\dfrac{q_z}{q_p}\right)^2 - x^2 \beta^2\left(\beta^2_{31}\dfrac{q_z}{q_p}\right)^2-y^2 \beta^2\left(\beta^2_{31}\dfrac{q_z}{q_p}\right)^2- 2 x^2 z \beta^2\left(\beta^2_{31}\dfrac{q_z}{q_p}\right)^2\nonumber\\
	&\qquad\qquad\qquad\qquad\qquad- z^2 \beta^2\left(\beta^2_{31}\dfrac{q_z}{q_p}\right)^2 - 
	2 x^2 z^2 \beta^2\left(\beta^2_{31}\dfrac{q_z}{q_p}\right)^2 - 2 y^2 z^2 \beta^2\left(\beta^2_{31}\dfrac{q_z}{q_p}\right)^2 + 
	z^2\left(\beta^2_{31}\dfrac{q_z}{q_p}\right)^4\nonumber\\ &\qquad\qquad\qquad\qquad\qquad+z^4\left(\beta^2_{31}\dfrac{q_z}{q_p}\right)^4+2 x\left(x^2 \beta^4 + y^2 \beta^4+ z (2 + 2 z - z \beta^2)\left(\beta^2_{31}\dfrac{q_z}{q_p}\right)^2\right) \cos\phi+
	4 x^2 z^2\left(\beta^2_{31}\dfrac{q_z}{q_p}\right)^2\nonumber\\
	&\qquad\qquad\qquad\qquad\qquad+2 y\Bigg(x^2 \beta^4 + y^2 \beta^4+z(2 + 2 z - z \beta^2)\left(\beta^2_{31}\dfrac{q_z}{q_p}\right)^2\Bigg)\sin\phi- 
	2 y^2 z \beta^2\left(\beta^2_{31}\dfrac{q_z}{q_p}\right)^2+2 z^3\left(\beta^2_{31}\dfrac{q_z}{q_p}\right)^4,\nonumber\\
	&\pi_6\left(x,y,z,\phi,\beta\right)=4 (x^2 + y^2) z^2 + z^4 - 
	2 (x^2 + y^2) z^2 \beta^2 + (x^2 + y^2)^2 \beta^4,\nonumber\\
	&\pi_7\left(x,y,z,\phi,\beta\right)=4 z^2 (x + \cos\phi)^2 + 
	4 z^2 (y +\sin\phi)^2 + (z^2 - \beta^2 (1 + x^2 + y^2 + 2 x \cos\phi + 2 y \sin\phi))^2.
\end{flalign}
Upon defining the terms $\xi^2_p=t^2_1q^2_p/(\beta^2_{31}|\Omega|)$, $\xi^2_z=t^2_3q^2_z/|\Omega|$ and $\xi^2_{zp}=\xi^2_z/\xi^2_p=\beta^4_{31}q^2_z/q^2_p$, we introduce the parametrization $\xi_p=\xi_R\cos\theta$ and $\xi_{zp}=\xi_R\sin\theta$ in terms of a positive-valued amplitude $\xi_R$ and azimuthial angle $\theta\in[0,2\pi]$ to re-write Eq.~\eqref{eq::LargeNfBosonicSelfEnergyRewritten} as
\begin{flalign}\label{eq::LargeNfBosonicSelfEnergyRewrittenXiParameters}
	\Pi(q_z,\xi_R,\theta,\phi,\beta,\beta_{31},t_1)&=-\dfrac{N_fe^2|q_z|}{(2\pi)^3t^2_1}\beta^2_{31}\int^\infty_{-\infty}dx\int^\infty_{-\infty}dy\int^\infty_{-\infty}dz\left[1+\dfrac{\pi_5\left(x,y,z,\phi,\beta,\xi_R\sin\theta\right)}{\sqrt{\pi_6\left(x,y,z,\phi,\beta\right)}\sqrt{\pi_7\left(x,y,z,\phi,\beta\right)}}\right]\times\nonumber\\
&\times\dfrac{\sqrt{\pi_6\left(x,y,z,\phi,\beta\right)}+\sqrt{\pi_7\left(x,y,z,\phi,\beta\right)}}{\left(\sqrt{\pi_6(\left(x,y,z,\phi,\beta\right)}+\sqrt{\pi_7\left(x,y,z,\phi,\beta\right)}\right)^2+\left(\dfrac{1}{\xi_R\cos\theta}\right)^2},\nonumber\\
\end{flalign}
We construct the ansatz for the expression in Eq.~\eqref{eq::LargeNfBosonicSelfEnergyRewritten} as a linear combination of the ansätze from Eqs.~\eqref{eq:AnsatzLargeNfBosonicSelfEnergyQp} and \eqref{eq:AnsatzLargeNfBosonicSelfEnergyQz}:
\begin{flalign}\label{eq:AnsatzLargeNfBosonicSelfEnergy}
	\widetilde{\Pi}(q_p,q_z,i\Omega,\beta,\beta_{31},t_1)&=-\dfrac{N_fe^2}{(2\pi)^3t^2_1}\sqrt{a^2_p(\beta)q^2_p+a^2_z(\beta)\beta^4_{31}q^2_z}\tanh(\sqrt{b^2_p(\beta)\dfrac{t^2_1}{\beta^2_{31}}\dfrac{q^2_p}{|\Omega|}+b^2_z(\beta)t^2_1\beta^2_{31}\dfrac{q^2_z}{|\Omega|}}).
\end{flalign}
Under the redefinitions $\xi^2_p=t^2_1q^2_p/(\beta^2_{31}|\Omega|)$, $\xi^2_z=t^2_3q^2_z/|\Omega|$ and $\xi^2_{zp}=\xi^2_3/\xi^2_p=\beta^4_{31}q^2_z/q^2_p$, we introduce the parametrization $\xi_p=\xi_R\cos\theta$ and $\xi_{zp}=\xi_R\sin\theta$, where $\xi_R\in[0,\infty)$ and $\theta\in[0,2\pi]$, to write Eq.~\eqref{eq:AnsatzLargeNfBosonicSelfEnergy} as
\begin{flalign}\label{eq:AnsatzLargeNfBosonicSelfEnergyXiParameters}
	\widetilde{\Pi}(q_z,\xi_R,\theta,\beta,\beta_{31},t_1)&=-\dfrac{N_fe^2|q_z|}{(2\pi)^3t^2_1}\beta^2_{31}\sqrt{\dfrac{a^2_p(\beta)}{\xi^2_R\sin^2\theta}+a^2_z(\beta)}\tanh(|\xi_R\cos\theta|\sqrt{b^2_p(\beta)+b^2_z(\beta)\xi^2_R\sin^2\theta}),
\end{flalign}
In Fig.~\ref{fig:AnsatzLargeNfBosonicSelfEnergyBeta}, the numerically integrated expression in Eq.~\eqref{eq::LargeNfBosonicSelfEnergyRewrittenXiParameters} is plotted against the ansatz given in Eq.~\eqref{eq:AnsatzLargeNfBosonicSelfEnergyXiParameters} in terms of the parameters $\{\xi_R,\theta,\beta\}$.
The agreement between numerical data and the continuous function is qualitatively satisfactory. We note, once again, that the numerically integrated result does not depend on the choice of the azimuthial angle $\phi$, as was the case for the ansatz in Eq.~\eqref{eq:AnsatzLargeNfBosonicSelfEnergyQp}. The renormalized bosonic Matsubara Green's function then writes
\begin{equation}
	D_{3,r}(\bold{q},i\Omega)=\left[D^{-1}(\bold{q},i\Omega)-	\Pi(\boldsymbol{q},i\Omega)\right]^{-1},
\end{equation}
where we have made the admission
\begin{equation}
	\Pi({\boldsymbol{q},i\Omega})=-\dfrac{N_fe^2}{(2\pi)^3t^2_1}\sqrt{a^2_p(\beta)q^2_p+a^2_z(\beta)\beta^4_{31}q^2_z}\tanh(\sqrt{b^2_p(\beta)\dfrac{t^2_1}{\beta^2_{31}}\dfrac{q^2_p}{|\Omega|}+b^2_z(\beta)t^2_1\beta^2_{31}\dfrac{q^2_z}{|\Omega|}}).
\end{equation}
To ease the notation, we henceforth refrain from writing the explicit dependence of the bosonic self-energy $\Pi({\boldsymbol{q},i\Omega})$ on the parameters $\{\beta,\beta_{31},t_1\}$. Upon considering the following result
\begin{equation}
	\lim_{\Omega\to0}\tanh(\sqrt{b^2_p(\beta)\dfrac{t^2_1}{\beta^2_{31}}\dfrac{q^2_p}{|\Omega|}+b^2_z(\beta)t^2_1\beta^2_{31}\dfrac{q^2_z}{|\Omega|}})=1,
\end{equation}
the static bosonic self-energy $\Pi(\boldsymbol{q},i\Omega\to i0)$ reads
\begin{flalign}
	\Pi(\boldsymbol{q},i\Omega\to i0)\to-\dfrac{N_fe^2}{(2\pi)^3t^2_1}\sqrt{a^2_p(\beta)q^2_p+a^2_z(\beta)\beta^4_{31}q^2_z}.
\end{flalign}
Furthermore, expanding the latter expression around the origin of momentum axes, we obtain the result
\begin{flalign}
	\Pi(\boldsymbol{q}\to\boldsymbol{0},i\Omega\to i0)\approx-\dfrac{N_fe^2}{(2\pi)^3t^2_1}\left(|a_p(\beta)|q_p+|a_z(\beta)|\beta^2_{31}|q_z|\right).
\end{flalign}
Therefore, the corresponding renormalized bosonic Matsubara Green's function becomes
\begin{flalign}
	D_{3,r}(\bold{q}\to0,i\Omega\to i0)&\approx\left[\dfrac{N_fe^2}{(2\pi)^3t^2_1}\left(|a_p(\beta)|q_p+|a_z(\beta)|\beta^2_{31}|q_z|\right)\right]^{-1}.
\end{flalign}
Thus, to leading-order in the momentum expansion, the latter expression is dominated by the contribution of the bosonic self-energy, leading the bosonic anomalous dimension to be $\eta^{(N_f\to\infty)}_\phi\sim0.5$. In the subsequent calculations, we proceed with the following approximate form
\begin{flalign}
	D_{3,r}(\bold{q},i\Omega)\approx\left[-\Pi(\boldsymbol{q},i\Omega)\right]^{-1}.
\end{flalign}
Finally, we note that the linear-in-momentum dependence of the bosonic self-energy $\Pi(\boldsymbol{q}\to\boldsymbol{0},i\Omega\to i0)$ implies the following real-space representation for the renormalized static Coulomb potential: 
\begin{equation}\label{eq:RenormalizedCoulombPotential}
	V(x,y,0)\propto\dfrac{1}{|\alpha_p(\beta)|\left(x^2+y^2\right)}\qquad\text{and}\qquad V(0,0,z)\propto\dfrac{1}{|\alpha_z(\beta)|\beta^2_{31}z^2}.
\end{equation}
Compared to the expressions $V_0(x,y,0)\propto\left[c^2_p\sqrt{x^2+y^2}\right]^{-1}$ and $V_0(0,0,z)\propto\left[c^2_z|z|\right]^{-1}$, the renormalized \textit{long-range} part of the Coulomb potential in Eq.~\eqref{eq:RenormalizedCoulombPotential} is universally suppressed (\textit{screened}).
\subsection{Fermionic self-energy}
The one-loop fermionic self-energy reads \cite{FetterWalecka}
\begin{flalign}\label{eq:LargeNfFermionicSelfEnergy}
	\dfrac{\Sigma(\bold{k},i\omega)}{e^2T}&=\sum_{i\Omega}\int\dfrac{d^3q}{(2\pi)^3}\,G_3(\bold{k}+\bold{q},i\omega+i\Omega)D_{3,r}(\bold{q},i\Omega)\nonumber\\
	&=\sum_{i\Omega}\int\dfrac{d^3q}{(2\pi)^3}\dfrac{i(\Omega+\omega)\mathbb{1}_2+\sum^3_{i=1}d_i(\bold{k}+\bold{q})\sigma_i}{(\Omega+\omega)^2+\sum^3_{i=1}d^2_i(\bold{k}+\bold{q})}\dfrac{1}{-\Pi(\boldsymbol{q},i\Omega)}.
\end{flalign}
\subsubsection{$\omega$-correction}
The correction to the fermionic Matsubara frequency is given by the expression
\begin{flalign}
	\delta_\omega=\dfrac{\text{Tr}\left(\mathbb{1}_2\frac{\partial}{-i\partial\omega}\Sigma(\bold{k},i\omega)\right)}{\text{Tr}\left(\mathbb{1}_2\right)}|_{\bold{k}=\bold{0},\,\omega=0}&=2\pi T\sum_{i\Omega}\int^\infty_0q_pdq_p\int^\infty_{-\infty}dq_z\dfrac{1}{N_ft^2_1}\dfrac{q^4_p\beta^4_{21}+q^4_z\beta^4_{31}+2q^2_pq^2_z(2-\beta^2_{21}\beta^2_{31})-\frac{\Omega^2}{t^4_1}}{\left[q^4_p\beta^4_{21}+q^4_z\beta^4_{31}+2q^2_pq^2_z(2-\beta^2_{21}\beta^2_{31})+\frac{\Omega^2}{t^4_1}\right]^2}\times\nonumber\\
	&\times\dfrac{\coth(\dfrac{t_1}{\sqrt{|\Omega|}}\sqrt{b^2_p(\beta_{21}\beta_{31})\dfrac{q^2_p}{\beta^2_{31}}+b^2_z(\beta_{21}\beta_{31})\beta^2_{31}q^2_z})}{\sqrt{a^2_p(\beta_{21}\beta_{31})q^2_p+a^2_z(\beta_{21}\beta_{31})\beta^4_{31}q^2_z}},
\end{flalign}
where we have substituted the expression $\beta=\beta_{21}\beta_{31}$ to make the dependence on the ratios of fermionic anisotropy parameters explicit. Through the variable transformations $X=q_p/\sqrt{\Omega/t^2_1}$ and $Y=q_z/\sqrt{\Omega/t^2_1}$, the above expression becomes
\begin{flalign}
	\delta_\omega=&\dfrac{4\pi}{N_f} T\sum_{i\Omega}\dfrac{1}{\Omega}\int^\infty_0XdX\int^\infty_0dY\dfrac{X^4\beta^2_{21}+Y^4\beta^4_{31}+2X^2Y^2(2-\beta^2_{21}\beta^2_{31})-1}{\left[X^4\beta^2_{21}+Y^4\beta^4_{31}+2X^2Y^2(2-\beta^2_{21}\beta^2_{31})+1\right]^2}\times\nonumber\\
	&\times\dfrac{\coth(\sqrt{b^2_p(\beta_{21}\beta_{31})\dfrac{X^2}{\beta^2_{31}}+b^2_z(\beta_{21}\beta_{31})\beta^2_{31}Y^2})}{\sqrt{a^2_p(\beta_{21}\beta_{31})X^2+a^2_z(\beta_{21}\beta_{31})\beta^4_{31}Y^2}},
\end{flalign}
where we used the fact that the integrand is invariant under the transformation $Y\to-Y$. The summation $T\sum_{i\Omega}\Omega^{-1}$ exhibits a singularity at zero frequency. Since we are interested in the zero-temperature behavior of the fermionic self-energy, we transform the discrete sum into an integral:
\begin{equation}
	T\sum_{i\Omega}\to\dfrac{1}{2\pi}\int^{\infty}_{-\infty}d\Omega.
\end{equation}
Therefore, upon introducing the chemical potential $\mu$ and parameter $\Lambda_0$ as the appropriate IR and UV cutoffs, respectively, the frequency integral is evaluated to be
\begin{equation}
	\dfrac{1}{2\pi}\int^{\infty}_{-\infty}\dfrac{d\Omega}{\Omega}=\dfrac{2}{2\pi}\int^{\infty}_{0}\dfrac{d\Omega}{\Omega}=\dfrac{2}{2\pi}\int^{\Lambda_0}_{\mu}\dfrac{d\Omega}{\Omega}=\dfrac{2}{2\pi}\text{ln}\,\Lambda_0/\mu.
\end{equation}
Finally, the correction to the fermionic Matsubara frequency reads
\begin{flalign}\label{eq:OmegaCorrection}
	\delta_\omega&=-\dfrac{\text{ln}\,\Lambda_0/\mu}{N_f}\Sigma_\omega(\beta_{21},\beta_{31}),
\end{flalign}
with
\begin{flalign}
	\Sigma_\omega(\beta_{21},\beta_{31})=&-\dfrac{4}{N_f}\int^\infty_0XdX\int^\infty_0dY\dfrac{X^4\beta^2_{21}+Y^4\beta^4_{31}+2X^2Y^2(2-\beta^2_{21}\beta^2_{31})-1}{\left[X^4\beta^2_{21}+Y^4\beta^4_{31}+2X^2Y^2(2-\beta^2_{21}\beta^2_{31})+1\right]^2}\times\nonumber\\
	&\times\dfrac{\coth(\sqrt{b^2_p(\beta_{21}\beta_{31})\dfrac{X^2}{\beta^2_{31}}+b^2_z(\beta_{21}\beta_{31})\beta^2_{31}Y^2})}{\sqrt{a^2_p(\beta_{21}\beta_{31})X^2+a^2_z(\beta_{21}\beta_{31})\beta^4_{31}Y^2}}.
\end{flalign}
\subsubsection{$t^2_1$-correction}
The correction to the fermionic anisotropy parameter $t^2_1$ is given by the expression
\begin{flalign}
	&\delta_{t^2_1}=\dfrac{\text{Tr}\left(\sigma_{1,2}\frac{\partial}{2t^2_1\partial k_{x,y}k_z}\Sigma(\bold{k},i\omega)\right)}{\text{Tr}\left(\sigma_{1,2}\right)}|_{\bold{k}=\bold{0},\,\omega=0}=8\pi T\sum_{i\Omega}\int^\infty_0q_pdq_p\int^\infty_{-\infty}dq_z\dfrac{1}{N_ft^2_1}\times\nonumber\\
	&\times\dfrac{q_p^8 \beta^8_{21} + 3 q^8_z \beta^8_{31} + 
		2 q^6_z q^2_p\beta^4_{31}(-2 + \beta^2_{21}\beta^2_{31})- \frac{\Omega^4}{t^8_1} - 
		6 q^2_z q^2_p(2 - \beta^2_{21} \beta^2_{31}) (q^4_p \beta^4_{21} - \frac{\Omega^2}{t^4_1}) + 
		2 q^4_z \beta^4_{31} (-6 q^4_p \beta^4_{21} +\frac{\Omega^2}{t^4_1})}{\left[q^4_p\beta^4_{21}+q^4_z\beta^4_{31}+2q^2_pq^2_z(2-\beta^2_{21}\beta^2_{31})+\frac{\Omega^2}{t^4_1}\right]^3}\times\nonumber\\
	&\times\dfrac{\coth(\dfrac{t_1}{\sqrt{|\Omega|}}\sqrt{b^2_p(\beta_{21}\beta_{31})\dfrac{q^2_p}{\beta^2_{31}}+b^2_z(\beta_{21}\beta_{31})\beta^2_{31}q^2_z})}{\sqrt{a^2_p(\beta_{21}\beta_{31})q^2_p+a^2_z(\beta_{21}\beta_{31})\beta^4_{31}q^2_z}}.
\end{flalign}
Upon following the computation steps described in the $\omega$-correction case, we find
\begin{flalign}\label{eq:t1^2Correction}
	\delta_{t^2_1}&=-\dfrac{\text{ln}\,\Lambda_0/\mu}{N_f}\Sigma_1(\beta_{21},\beta_{31}),
\end{flalign}
with
\begin{flalign}
	&\Sigma_1(\beta_{21},\beta_{31})=-\dfrac{4}{N_f}\int^\infty_0XdX\int^\infty_0dY\,\dfrac{\coth(\sqrt{b^2_p(\beta_{21}\beta_{31})\dfrac{X^2}{\beta^2_{31}}+b^2_z(\beta_{21}\beta_{31})\beta^2_{31}Y^2})}{\sqrt{a^2_p(\beta_{21}\beta_{31})X^2+a^2_z(\beta_{21}\beta_{31})\beta^4_{31}Y^2}}\times\nonumber\\
	&\times\dfrac{X^8 \beta^8_{21} + 3 Y^8 \beta^8_{31} + 
		2 Y^6 X^2\beta^4_{31}(-2 + \beta^2_{21}\beta^2_{31})-1- 
		6 Y^2 X^2(2 - \beta^2_{21} \beta^2_{31}) (X^4 \beta^4_{21} -1) + 
		2 Y^4 \beta^4_{31} (-6X^4 \beta^4_{21} +1)}{\left[X^4\beta^2_{21}+Y^4\beta^4_{31}+2X^2Y^2(2-\beta^2_{21}\beta^2_{31})+1\right]^3}.
\end{flalign}
\subsubsection{$t^2_2$-correction}
Upon following the computation steps described in the $\omega$-correction case, the correction to the fermionic anisotropy parameter $t^2_2$ reads
\begin{flalign}\label{eq:t2^2Correction}
	\delta_{t^2_2}=\dfrac{\text{Tr}\left(\sigma_3\frac{\partial}{t^2_2\partial k_{x,y}k_{x,y}}\Sigma(\bold{k},i\omega)\right)}{\text{Tr}\left(\sigma_3\right)}|_{\bold{k}=\bold{0},\,\omega=0}&=-\dfrac{\text{ln}\,\Lambda_0/\mu}{N_f}\Sigma_2(\beta_{21},\beta_{31}),
\end{flalign}
with
\begin{flalign}
	&\Sigma_2(\beta_{21},\beta_{31})=8\int^\infty_0XdX\int^\infty_0dY\,\dfrac{\coth(\sqrt{b^2_p(\beta_{21}\beta_{31})\dfrac{X^2}{\beta^2_{31}}+b^2_z(\beta_{21}\beta_{31})\beta^2_{31}Y^2})}{\sqrt{a^2_p(\beta_{21}\beta_{31})X^2+a^2_z(\beta_{21}\beta_{31})\beta^4_{31}Y^2}}\times\nonumber\\
	&\times\left[X^8 \beta_{21}^{10} - 12 Y^4 X^4 \beta_{21}^4 \beta_{31}^2 + 
		Y^8 \beta_{31}^6 (4 - \beta_{21}^2 \beta_{31}^2) - 
		2 Y^2 X^6 \beta_{21}^6 (2 + \beta_{21}^2 \beta_{31}^2) + 
		2 Y^6 X^2 \beta_{31}^2 (-8 + 
		6 \beta_{21}^2 \beta_{31}^2 + \beta_{21}^4 \beta_{31}^4)\right.\nonumber\\
	&\left.+2 (2 Y^2 + 3 X^2 \beta_{21}^4) (-X \beta_{21} + 
	 	Y \beta_{31})(X \beta_{21} + Y \beta_{31}) + \beta_{21}^2\right]\dfrac{1}{\left[X^4\beta^2_{21}+Y^4\beta^4_{31}+2X^2Y^2(2-\beta^2_{21}\beta^2_{31})+1\right]^3}.
\end{flalign}
\subsubsection{$t^2_3$-correction}
Upon following the computation steps described in the $\omega$-correction case, the correction to the fermionic anisotropy parameter $t^2_3$ reads
\begin{flalign}\label{eq:t3^2Correction}
	\delta_{t^2_3}=\dfrac{\text{Tr}\left(\sigma_3\frac{\partial}{-t^2_3\partial k_zk_z}\Sigma(\bold{k},i\omega)\right)}{\text{Tr}\left(\sigma_3\right)}|_{\bold{k}=\bold{0},\,\omega=0}&=-\dfrac{\text{ln}\,\Lambda_0/\mu}{N_f}\Sigma_3(\beta_{21},\beta_{31}),
\end{flalign}
with
\begin{flalign}
	&\Sigma_3(\beta_{21},\beta_{31})=8\int^\infty_0XdX\int^\infty_0dY\,\dfrac{\coth(\sqrt{b^2_p(\beta_{21}\beta_{31})\dfrac{X^2}{\beta^2_{31}}+b^2_z(\beta_{21}\beta_{31})\beta^2_{31}Y^2})}{\sqrt{a^2_p(\beta_{21}\beta_{31})X^2+a^2_z(\beta_{21}\beta_{31})\beta^4_{31}Y^2}}\times\nonumber\\
	&\times\left[3Y^8 \beta_{31}^{10}+ X^8 \beta_{21}^6 (4 - \beta_{21}^2 \
	\beta_{31}^2) + 
	6 Y^4 X^4 \beta_{21}^2 \beta_{31}^4 (-6 + \beta_{21}^2 \
	\beta_{31}^2) - 
	4 Y^6X^2 \beta_{31}^6 (1 + 2 \beta_{21}^2 \beta_{31}^2) + 
	12 Y^2X^6 \beta_{21}^2 (-4 + 3 \beta_{21}^2 \beta_{31}^2)\right.\nonumber\\
	&\left. + 
	4 (X^4 \beta_{21}^2 - 3 Y^4 \beta_{31}^6 + 
	3 Y^2X^2 \beta_{31}^2 (-1 + \beta_{21}^2 \beta_{31}^2)) + \
	\beta_{31}^2\right]\dfrac{1}{\left[X^4\beta^2_{21}+Y^4\beta^4_{31}+2X^2Y^2(2-\beta^2_{21}\beta^2_{31})+1\right]^3}.
\end{flalign}
\subsubsection{$e$-correction}
Finally, the correction to the electronic charge $e$ reads \cite{FetterWalecka}
\begin{flalign}
	\delta_{e}&=e^2T\sum_{i\Omega}\int\dfrac{d^3q}{(2\pi)^3}\dfrac{\text{Tr}\left[G_3(\bold{q},i\Omega)G_3(\bold{q},i\Omega)\right]}{2}D_{3,r}(\bold{q},i\Omega)\nonumber\\
	&=2\pi T\sum_{i\Omega}\int^\infty_0q_pdq_p\int^\infty_{-\infty}dq_z\dfrac{1}{N_ft^2_1}\dfrac{q^4_p\beta^4_{21}+q^4_z\beta^4_{31}-2q^2_zq^2_p(-2+\beta^2_{21}\beta^2_{31})-\frac{\Omega^2}{t^4_1}}{\left[q^4_p\beta^4_{21}+q^4_z\beta^4_{31}+2q^2_pq^2_z(2-\beta^2_{21}\beta^2_{31})+\frac{\Omega^2}{t^4_1}\right]^2}\times\nonumber\\
	&\times\dfrac{\coth(\dfrac{t_1}{\sqrt{|\Omega|}}\sqrt{b^2_p(\beta_{21}\beta_{31})\dfrac{q^2_p}{\beta^2_{31}}+b^2_z(\beta_{21}\beta_{31})\beta^2_{31}q^2_z})}{\sqrt{a^2_p(\beta_{21}\beta_{31})q^2_p+a^2_z(\beta_{21}\beta_{31})\beta^4_{31}q^2_z}}=\delta_\omega,
\end{flalign}
which is identical to the correction for the fermionic Matsubara frequency. The latter result is a proof of the Ward-Takahashi identity \cite{Ward1950,Takahashi1957} stating that the charge renormalization is equal to the frequency renormalization in order for the corresponding fermionic action to be the gauge invariant \cite{Mahan1990}.
\subsection{Emergent spatial anisotropy and breakdown of Fermi-liquid behavior}
Considering the renormalization within the large-$N_f$ scheme (Eqs.~\eqref{eq:OmegaCorrection},\eqref{eq:t1^2Correction},\eqref{eq:t2^2Correction},\eqref{eq:t3^2Correction}), the fermionic self-energy reads
\begin{flalign}\label{eq:FermionicSelfEnergyNfLarge}
	\Sigma(\bold{k},i\omega)=&-\left[-\dfrac{\text{ln}\,\Lambda_0/\mu}{N_f}\Sigma_\omega(\beta_{21},\beta_{31})\right] i\omega\mathbb{1}_2+\left[-\dfrac{\text{ln}\,\Lambda_0/\mu}{N_f}\Sigma_1(\beta_{21},\beta_{31})\right]2t^2_1k_z(k_x\sigma_1+k_y\sigma_2)\nonumber\\
	&+\left[-\dfrac{\text{ln}\,\Lambda_0/\mu}{N_f}\Sigma_2(\beta_{21},\beta_{31})\right]t^2_2(k^2_x+k^2_y)\sigma_3-\left[-\dfrac{\text{ln}\,\Lambda_0/\mu}{N_f}\Sigma_3(\beta_{21},\beta_{31})\right]t^2_3k^2_z\sigma_3
\end{flalign}
and the renormalized fermionic action involving the Hamiltonian given in Eq.~\eqref{eq:QuadraticHamiltonian} writes
\begin{flalign}\label{eq:RenormalizedFermionicActionLargeNf}
	&\mathcal{S}_f=T\sum_{i\omega}\int\dfrac{d^3k}{(2\pi)^3}\,\psi^\dag(\bold{k},i\omega)\left[-i\omega\mathbb{1}_2+H(\bold{k})-\Sigma(\bold{k},i\omega)\right]\psi(\bold{k},i\omega).
\end{flalign}
Since the expression in Eq.~\eqref{eq:FermionicSelfEnergyNfLarge} depends on the ratios of fermionic anisotropy parameters $\{\beta^2_{21},\beta^2_{31}\}$, we treat the ratio $\ell=\Lambda_0/\mu$ as the dimensionless parameter initiating scale transformations. Then, we construct the following equations describing the evolution of the parameters $\{\beta^2_{21},\beta^2_{31}\}$ as functions of the scaling parameter $\ell$:
 \begin{figure}[t]
	\centering
	\subfloat{\includegraphics[width=0.35\textwidth]{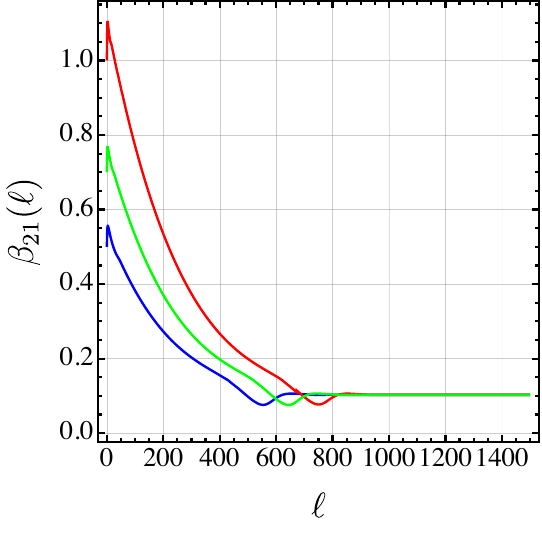}}\quad
	\subfloat{\includegraphics[width=0.35\textwidth]{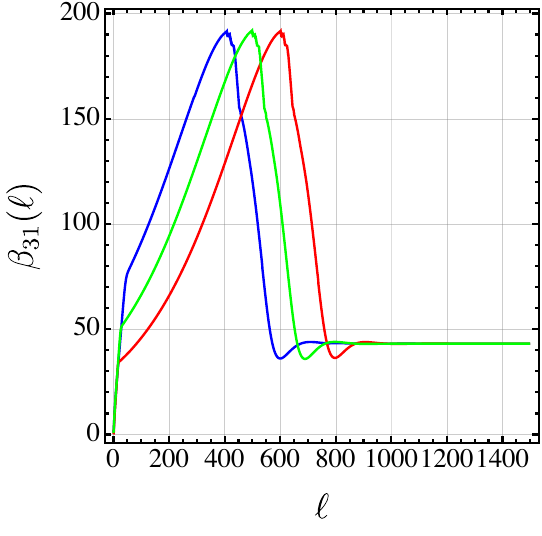}}
	\caption{Evolution of the ratios of fermionic anisotropy parameters $\beta_{21}$ and $\beta_{31}$ according to Eq.~\eqref{eq:RGEquationsLargeNfLimit} in terms of the dimensionless scaling parameter $\ell$. The color code in both plots represent the initial values for the aforementioned two parameters, $(\beta^{(0)}_{21},\beta^{(0)}_{31})$: $(0.5,0.8)$ (blue), $(1,0.1)$ (red), $(0.7,0.7)$ (green).}
	\label{fig:LargeNfFlowFermionicAnisotropyRatios}
\end{figure}
\begin{flalign}\label{eq:RGEquationsLargeNfLimit}
	\dfrac{1}{\beta^2_{21,31}}\dfrac{d\beta^2_{21,31}}{d\text{ln}\ell}=\Sigma_{2,3}(\beta_{21},\beta_{31})-\Sigma_1(\beta_{21},\beta_{31}).
\end{flalign}
As depicted in Fig.~\ref{fig:LargeNfFlowFermionicAnisotropyRatios}, there is a saturation for the estimated value of the ratios $\{\beta^2_{21},\beta^2_{31}\}$ in the limit of infinite scale transformations leading to the fixed-point solutions ${\beta^2_{21}}^*\approx0.0106$ and ${\beta^2_{31}}^*\approx1860$. Qualitatively, the latter result reads
\begin{equation}\label{eq:LargeNfFixedPoint}
	{\beta^2_{21}}^*\ll1\qquad\text{and}\qquad{\beta^2_{31}}^*\gg1,
\end{equation}
indicating that the fermionic anisotropy parameter $t^2_2$ is reduced to an infinitesimal value whereas $t^2_3$ becomes extremely large and $t^2_1$ remains non-zero but finite. Therefore, the screening due to the enhanced Coulomb potential (see Eq.~\eqref{eq:RenormalizedCoulombPotential}) yields a different renormalization for the three form factors of the fermionic Hamiltonian given in Eq.~\eqref{eq:QuadraticHamiltonian}, thus to an  \textit{emerging} spatial anisotropy. That feature has consequences also for the fermionic Matsubara Green's function. From Eq.~\eqref{eq:RenormalizedFermionicActionLargeNf}, one may define the fermionic anomalous dimension as \cite{Sachdev2011}
\begin{equation}\label{eq:LargeNfFermionicAnomalousDimension}
	\eta^{(N_f\to\infty)}_\psi=\dfrac{\Sigma_\omega(\beta_{21},\beta_{31})}{2N_f}
\end{equation}
which admits the fixed-point value ${\eta^{(N_f\to\infty)}_\psi}^*\approx0.034/N_f$. Hence, qualitatively, we can express the renormalized fermionic retarded Green's function as \cite{Sachdev2011}
\begin{equation}
	G^R_3(\bold{k},\omega+i0^+)=\dfrac{P_+(\hat{\bold{k}})}{2}\left[\dfrac{1}{\omega+i0^+-E_3(\bold{k})}\right]^{1-{\eta^{(N_f\to\infty)}_\psi}^*}+\dfrac{P_-(\hat{\bold{k}})}{2}\left[\dfrac{1}{\omega+i0^++E_3(\bold{k})}\right]^{1-{\eta^{(N_f\to\infty)}_\psi}^*},
\end{equation}
where $P_{\pm}(\hat{\bold{k}})=\mathbb{1_2}\pm f(\hat{\bold{k}},\sigma_x,\sigma_y,\sigma_z)$ are projection operators to the conduction and valence bands with the function $f(\hat{\bold{k}},\sigma_x,\sigma_y,\sigma_z)$ depending linearly on the involved Pauli matrices. We note that the fermionic spectral density function $S(\bold{k},\omega)=-(1/\pi)\text{Tr}\left[\text{Im}\,G^R_3(\bold{k},\omega+i0^+)\right]$ does \textit{not} exhibit sharp poles at the real frequencies $\omega=\pm E_3(\bold{k})$, but power-law singularities. Therefore, its form consists of Lorentzian distributions around the aforementioned poles. Such an outcome is an indication that the enhanced Coulomb interaction at three spatial dimensions leads to the \textit{breakdown} of the Fermi-liquid behavior \cite{Lohneysen2007}.
Finally, considering the fixed-point solution in Eq.~\eqref{eq:LargeNfFixedPoint} for the renormalized Coulomb potential given in Eq.~\eqref{eq:RenormalizedCoulombPotential}, we find the latter to assume the form
\begin{flalign}\label{eq:RenormalizedCoulombPotentialFixedPoint}
	&V(x,y,0)|_{{\beta^2_{21}}^*\ll1,\,{\beta^2_{31}\gg1}^*}\propto\dfrac{1}{\varepsilon_p(x^2+y^2)},\nonumber\\
	&V(0,0,z)|_{{\beta^2_{21}}^*\ll1,\,{\beta^2_{31}\gg1}^*}\propto\dfrac{1}{\varepsilon_zz^2},
\end{flalign}
where the \textit{effective} permittivities satisfy the relation $\varepsilon_p\ll1\ll\varepsilon_z$. Thus, the renormalized Coulomb potential is significantly suppressed along the rotational $\hat{z}$ axis compared to the other coordination axes implying that the interacting semimetallic phase at three dimensions exhibits strong spatial anisotropy.
\section{Details of the $\epsilon$-expansion calculation}
In the present section, we provide details regarding the renormalization-group analysis under the $\epsilon$-expansion method discussed in the corresponding section of the main text. For that purpose, the one-dimensional $\hat{z}$ axis is extended to a $d$-dimensional manifold by introducing the real-valued dimension $\epsilon=2-d$ lying in the interval $[0,1]$. The renormalization-group scheme is applied along the radius of the extended $d$-dimensional manifold and the corresponding angular content is considered in two dimensions. Such a choice naturally generates the artificial fermionic (bosonic) momentum $k_4$ $(q_4)$. Therefore, the fermionic and bosonic Hamiltonians given in Eqs.~\eqref{eq:QuadraticHamiltonian} and \eqref{eq:BosonicHamiltonian} obtain the four-dimensional forms
\begin{align}\label{eq:QuadraticDHamiltonian}
	H_4(\bold{k})=\sum^5_{i=1}d_i(\bold{k})\Gamma_i,
\end{align}
and
\begin{align}\label{eq:BosonicDHamiltonian}
	H_{4,b}(\bold{q})=c^2_p(q^2_x+q^2_y)+c^2_z(q^2_z+q^2_4),
\end{align}
with the form factors $d_1(\bold{k})=2t^2_1k_xk_z$, $d_2(\bold{k})=2t^2_1k_xk_4$, $d_3(\bold{k})=2t^2_1k_yk_z$, $d_4(\bold{k})=2t^2_1k_yk_4$ and $d_5(\bold{k})=t^2_2(k^2_x+k^2_y)-t^2_3(k^2_z+k^2_4)$. For the Hamiltonian $H_4(\bold{k})$ to accommodate four-dimensional form factors, the representation of the Clifford algebra has been extended from two to four dimensions using the Dirac matrices $\{\mathbb{1}_4,\dots,\Gamma_{14}\}$ respecting the fundamental relations $\{\Gamma_i,\Gamma_j\}=4\delta_{ij}\mathbb{1}_4$ and $\text{Tr}\qty[\Gamma_i\Gamma_j]=4\delta_{ij}$ with $i,j=\{1,\dots,14\}$; knowledge of the latter two properties of Dirac matrices is sufficient for the computation of one-loop diagrams involved in the renormalization-group analysis \cite{Snygg1997}. The corresponding fermionic and bosonic Matsubara Green's functions read, respectively,
\begin{flalign}
	G_4(\bold{k},i\omega)=\left[-i\omega\mathbb{1}_4+H_4(\bold{k})\right]^{-1},\qquad D_4(\bold{q},i\Omega)=\left[H_{4,b}(\bold{k})\right]^{-1}.
\end{flalign}
\subsection{Bosonic self-energy}
The one-loop \textit{three-dimensional} static bosonic self-energy is computed from the expression \cite{Sachdev2011}
\begin{flalign}\label{eq:BosonicSelfEnergy}
	\Pi(\bold{q})&=N_fe^2T\sum_{i\omega}\int^\Lambda_{\Lambda/\ell}\dfrac{d^4k}{(2\pi)^4}\Tr\left[G_4(\bold{k}+\bold{q},i\omega)G_4(\bold{k},i\omega)\right]\nonumber\\
	&=-N_fe^2\int^\Lambda_{\Lambda/\ell}\dfrac{d^4k}{(2\pi)^4}\dfrac{1}{E_4(\bold{k}+\bold{q})+E_4(\bold{k})}\left[1-\dfrac{\sum^5_{i=1}d_i(\bold{k}+\bold{q})d_i(\bold{k})}{E_4(\bold{k}+\bold{q})E_4(\bold{k})}\right].
\end{flalign}
where $E_4(\bold{k})=\sqrt{\sum^5_{i=1}d_i(\bold{k})}$ is the fermionic dispersion relation. The fermionic momenta lying in the radial interval $[\Lambda/\ell,\Lambda]$ of the $d$-dimensional manifold, with $\ell$ being the dimensionless scaling parameter and $\Lambda$ the momentum UV cutoff, are to be integrated upon. Our choice to estimate the three-dimensional bosonic self-energy implies the following summation rule among fermionic and bosonic momenta
\begin{equation}
	\bold{k}+\bold{q}=(k_x+q_x,k_y+q_y,k_z+q_z,k_4).
\end{equation}
The thus applied dimensional regularization scheme along the rotational $\hat{z}$ axis enables to extract the logarithmic contribution from the involved one-loop diagrams for the physically relevant case of a three-dimensional long-range Coulomb interaction.
\subsubsection{Projection to $q^2_{x,y}$ bosonic momenta}
The projected bosonic self-energy to momenta $q^2_{x,y}$ reads
\begin{flalign}
	\dfrac{\partial^2\Pi(\bold{q})}{\partial q^2_{x,y}}|_{q_z=0}=-\dfrac{N_fe^2q^2_{x,y}}{(2\pi)^3}\int^\infty_0k_pdk_p\int^\Lambda_{\Lambda/\ell}K^{d-1}dK\dfrac{K^2t^4_1\left(2k^2K^2t^4_1+k^4_pt^4_2+K^4t^4_3\right)}{\left[k^4_pt^4_2+K^4t^4_3+2k^2_pK^2(2t^4_1-t^2_2t^2_3)\right]^{5/2}}.
\end{flalign}
To obtain the latter expression, we first introduced the circular coordinates $k_x=k_p\cos\phi$ and $k_y=k_p\sin\phi$ with angle $\phi\in[0,2\pi]$ and radius $k_p\in[0,\infty)$ for the fermionic momenta perpendicular to the rotational $\hat{z}$ axis, and the circular coordinates $k_z=K\cos\theta$ and $k_4=K\sin\theta$ with angle $\theta\in[0,2\pi]$ and radius $K\in[\Lambda/\ell,\Lambda]$ for the extended $d$-dimensional manifold. Then, we performed the integrations over the angles $\{\phi,\theta\}$. Subsequently, we use the ratios of fermionic anisotropy parameters $\beta_{{21,3}1}=t_{2,3}/t_1$ as well as introduce the integration variable $r=k_p/K$ to perform the integration over the variable $K$:
\begin{flalign}
	\dfrac{\partial^2\Pi(\bold{q})}{\partial q^2_{x,y}}|_{q_z=0}=-\dfrac{N_fe^2q^2_{x,y}\Lambda^{-\epsilon}}{(2\pi)^3t^2_1}\text{ln}\ell\int^\infty_0rdr\,\dfrac{2\pi(2r^2+\beta^4_{21}r^4+\beta^4_{31})}{\left[4r^2+(r^2\beta^2_{21}-\beta^2_{31})^2\right]^{5/2}}.
\end{flalign}
The integral in the latter expression yields
\begin{flalign}
	\int^\infty_0rdr\,\dfrac{2\pi(2r^2+\beta^4_{21}r^4+\beta^4_{31})}{\left[4r^2+(r^2\beta^2_{21}-\beta^2_{31})^2\right]^{5/2}}=\dfrac{\pi\left(1+\dfrac{\beta^2_{21}\beta^2_{31}}{2}\right)}{3\beta^2_{31}}.
\end{flalign}
Therefore, we write
\begin{flalign}\label{eq:ProjectionBosonicMomentaQxQy}
	\dfrac{\partial^2\Pi(\bold{q})}{\partial q^2_{x,y}}|_{q_{z,4}=0}=-N_f\,\alpha\,\text{ln}\ell\,B(\gamma_{21},\gamma_{31}) c^2_p q^2_{x,y},
\end{flalign}
where we have introduced the function
\begin{equation}
	B(\gamma_{21},\gamma_{31}) =\dfrac{\gamma_{21}}{2}+\dfrac{1}{\gamma_{31}},
\end{equation}
with the dimensionless parameters
\begin{equation}
	\gamma_{21}=\dfrac{c_z}{c_p}\beta^2_{21},\qquad\gamma_{31}=\dfrac{c_p}{c_z}\beta^2_{31},
\end{equation}
as well as the (dimensionless) coupling constant
\begin{flalign}\label{eq:CouplingConstant}
	\alpha=\dfrac{e^2\Lambda^{-\epsilon}}{24\pi^2t^2_1c_pc_z}.
\end{flalign}
\subsubsection{Projection to $q^2_z$ bosonic momentum}
The projected bosonic self-energy to momentum $q^2_z$ reads
\begin{flalign}
	\dfrac{\partial^2\Pi(\bold{q})}{\partial q^2_z}|_{q_{x,y}=0}=-\dfrac{N_fe^2q^2_z}{(2\pi)^3}\int^\infty_0k_pdk_p\int^\Lambda_{\Lambda/\ell}K^{d-1}dK\,\dfrac{2k^2_p\pi t^4_1\left(2k^2_pK^2t^4_1+k^4_pt^4_2+K^4t^4_3\right)}{\left[k^4_pt^4_2+(k^2_z+k^2_4)^2t^2_3+2k^2_p(k^2_z+k^2_4)(2t^2_1-t^2_2t^2_3)\right]^{5/2}},
\end{flalign}
where we performed the computation steps described in the case for the projection to momenta $\{q^2_x,q^2_y\}$. Then, upon introducing the integration variable $r=k_p/K$, we perform the integration concerning the $d$-dimensional manifold to find
\begin{flalign}
	\dfrac{\partial^2\Pi(\bold{q})}{\partial q^2_z}|_{q_{x,y}=0}=-\dfrac{N_fe^2q^2_{z,4}\Lambda^{-\epsilon}}{(2\pi)^3t^2_1}\text{ln}\ell\int^\infty_0rdr\,\dfrac{2\pi r^2\left(2r^2+r^4\beta^4_{21}+\beta^4_{31}\right)}{\left[4r^2+(r^2\beta^2_{21}-\beta^2_{31})^2\right]^{5/2}},
\end{flalign}
where ratios of fermionic anisotropy parameters $\beta_{{2,3}1}=t_{2,3}/t_1$ were introduced. The remaining integral can be analytically computed to yield
\begin{equation}
	\int^\infty_0rdr\,\dfrac{2\pi r^2(r^2\beta^2_{21}+\beta^2_{31})^2}{\left[4r^2+(r^2\beta^2_{21}-\beta^2_{31})^2\right]^{5/2}}=\dfrac{\pi\left(4+2\beta^2_{21}\beta^2_{31}\right)}{12\beta^2_{21}}.
\end{equation}
Thus, we arrive at the final expression
\begin{flalign}\label{eq:ProjectionBosonicMomentaQzQ4}
	\dfrac{\partial^2\Pi(\bold{q})}{\partial q^2_z}|_{q_{x,y}=0}=-N_f\,\alpha\,\text{ln}\ell\,B(\gamma_{31},\gamma_{21}) c^2_z q^2_z.
	\end{flalign}
\subsubsection{Renormalization-group equations from the $\epsilon$-expansion}
Collecting the contributions from Eqs.~\eqref{eq:ProjectionBosonicMomentaQxQy} and \eqref{eq:ProjectionBosonicMomentaQzQ4}, the bosonic self-energy reads
\begin{flalign}
	\Pi(\bold{q})=-N_f\,\alpha\,\text{ln}\ell\,B(\gamma_{21},\gamma_{31})c^2_p(q^2_x+q^2_y)-N_f\,\alpha\,\text{ln}\ell\,B(\gamma_{31},\gamma_{21})c^2_zq^2_z.
\end{flalign}
Then, the three-dimensional renormalized bosonic action writes
\begin{flalign}\label{eq:BosonicAction}
	S_b=\dfrac{1}{2}T\sum_{i\Omega}\int\dfrac{d^3q}{(2\pi)^3}\phi^\dag(\bold{q},i\Omega)\left[H_{b}(\bold{q})-\Pi(\bold{q})\right]\phi(\bold{q},i\Omega).
\end{flalign}
Considering the characteristic length scale $L$, Eq.~\eqref{eq:BosonicAction} can be made \textit{self-similar} to the bare bosonic action upon introducing the dimensionless quantities
\begin{flalign}\label{eq:DimensionlessQuantities1}
	\widetilde{\phi}^{(\dag)}=\phi^{(\dag)}L^{-\frac{7}{2}},\qquad\widetilde{\Omega}=\Omega L^2,\qquad\widetilde{q}=qL,\qquad\widetilde{T}=TL^2,\qquad\widetilde{c}^2_{p,z}=c^2_{p,z}L^0,\qquad\widetilde{e}^2=e^2L^{\epsilon}.
\end{flalign}
Then, the one-loop renormalization-group (RG) equations for the dimensionless bosonic anisotropy parameters $\widetilde{c}^2_{p,z}$ are
\begin{flalign}\label{eq:BosonicRGEquations}
	\dfrac{1}{\widetilde{c}^2_p}\dfrac{d\widetilde{c}^2_p}{d\text{ln}\ell}=N_f\,\alpha\,B(\gamma_{21},\gamma_{31}),\qquad\qquad\dfrac{1}{\widetilde{c}^2_z}\dfrac{d\widetilde{c}^2_z}{d\text{ln}\ell}=N_f\,\alpha\,B(\gamma_{31},\gamma_{21}).
\end{flalign}
\subsection{Fermionic self-energy}
The one-loop \textit{three-dimensional} static fermionic self-energy is computed from the expression \cite{Sachdev2011}
\begin{flalign}\label{eq:FermionicSelfEnergy}
	\Sigma(\bold{k})&=e^2T\sum_{i\Omega}\int^\Lambda_{\Lambda/\ell}\dfrac{d^4q}{(2\pi)^4}G_3(\bold{k}+\bold{q},i\Omega)D_4(\bold{q},i\Omega)\nonumber\\
	&=\sum_{i\Omega}\int^\Lambda_{\Lambda/\ell}\dfrac{d^4q}{(2\pi)^4}\dfrac{i\Omega\mathbb{1}_2+\sum^3_{i=1}d_i(\bold{k}+\bold{q})\sigma_i}{\Omega^2+\sum^3_{i=1}d^2_i(\bold{k}+\bold{q})}\dfrac{1}{H_{4,b}(\bold{q},i\Omega)}.
\end{flalign}
The bosonic momenta to be integrated upon lie in the radial interval $[\Lambda/\ell,\Lambda]$ of the extended $d$-dimensional manifold. Our choice to estimate the three-dimensional fermionic self-energy implies the following summation rule among fermionic and bosonic momenta
\begin{equation}
	\bold{k}+\bold{q}=\left(k_x+q_x,k_y+q_y,k_z+\sqrt{q^2_z+q^2_4}\right).
\end{equation}
In the spirit of the calculation for the bosonic self-energy, the dimensional regularization scheme along the rotational $\hat{z}$ axis of the bosonic momenta enables to extract the logarithmic contribution from the involved one-loop diagrams for the physically relevant case of a three-dimensional quadratic semimetal.
\subsubsection{Projection to $k_{x,y}k_z$ fermionic momenta}
The projected fermionic self-energy to momenta $k_{x,y}k_z$ reads
\begin{flalign}
	\dfrac{\text{Tr}\left(\sigma_{1,2}\frac{\partial^2\Sigma(\bold{k})}{\partial k_{x,y}k_z}\right)}{2t^2_1\text{Tr}\left(\sigma_{1,2}\right)}|_{k_{x,y,z}=0}=&-\dfrac{e^2}{2(2\pi)^3}\int^\infty_0q_pdq_p\int^\Lambda_{\Lambda/\ell}Q^{d-1}dQ\,\Bigg[6q^2_pQ^4t^4_2-3q^6_pQ^2t^6_2t^2_3+2q^2_pQ^4t^4_3+5q^4_pQ^2t^4_2t^4_3\Bigg.\nonumber\\
	&\Bigg.-q^2_pQ^4t^2_2t^6_3-Q^6t^8_3\Bigg]\dfrac{4\pi Q^2t^2_1}{\left(c^2_pq^2_p+c^2_zQ^2\right)\left[q^4_pt^4_2+Q^4t^4_3+2q^2_pQ^2(2t^4_1-t^2_2t^2_3)\right]^{5/2}}.
\end{flalign}
To obtain the latter expression, we first introduced the circular coordinates $q_x=q_p\cos\phi$ and $q_y=q_p\sin\phi$ with angle $\phi\in[0,2\pi]$ and radius $q_p\in[0,\infty)$ for the bosonic momenta perpendicular to the rotational $\hat{z}$ axis, and the circular coordinates $q_z=Q\cos\theta$ and $q_4=Q\sin\theta$ with angle $\theta\in[0,2\pi]$ and radius $Q\in[\Lambda/\ell,\Lambda]$ for the extended $d$-dimensional manifold. Then, we performed the integrations over the angles $\{\phi,\theta\}$. Subsequently, we use the ratios of fermionic and bosonic anisotropy parameters $\gamma_{21}=c_z\beta^2_{21}/c_p$ and $\gamma_{31}=c_p\beta^2_{31}/c_z$ as well as introduce the variable $r=q_p/Q$ to perform the integration over $Q$
\begin{flalign}\label{eq:ProjectionFermionicMomentaKxyKz}
	\dfrac{\text{Tr}\left(\sigma_{1,2}\frac{\partial^2}{\partial k_{x,y}k_z}\Sigma(\bold{k})\right)}{2t^2_1\text{Tr}\left(\sigma_{1,2}\right)}|_{k_{x,y,z}=0}=-\alpha\,F_1(\gamma_{21},\gamma_{31})\,\text{ln}\ell,
\end{flalign}
where we have introduced the function
\begin{equation}
	F_1(\gamma_{21},\gamma_{31}) =\dfrac{3}{2\pi}\int^\infty_0rdr\,\dfrac{4\pi\left(6r^6\gamma^2_{21}-3r^6\gamma^3_{21}\gamma_{31}+(2r^2+5r^4\gamma^2_{21})\gamma^2_{31}-r^2\gamma_{21}\gamma^3_{31}-\gamma^4_{31}\right)}{\left(r^2+1\right)\left[4r^2+(r^2\gamma^2_{21}-\gamma^2_{31})^2\right]^{5/2}}.
\end{equation}
\subsubsection{Projection to $k^2_{x,y}$ fermionic momenta}
The projected fermionic self-energy to momenta $k^2_{x,y}$ reads
\begin{flalign}
	\dfrac{\text{Tr}\left(\sigma_3\frac{\partial^2\Sigma(\bold{k})}{\partial k^2_{x,y}}\right)}{t^2_2\text{Tr}\left(\sigma_3\right)}|_{k_{x,y,z}=0}=-\dfrac{e^2}{2(2\pi)^3}\int^\infty_0qdq\int^\Lambda_{\Lambda/\ell}Q^{d-1}dQ\,\dfrac{8\pi Q^2t^4_1\left(Q^2t^2_3-q^2t^2_2\right)\left(q^4t^4_2+Q^4t^4_3-2q^2Q^2(t^4_1-2t^2_2t^2_3)\right)}{\left(c^2_pq^2+c^2_zQ^2\right)\left[q^4t^4_2+Q^4t^4_3+2q^2Q^2(2t^4_1-t^2_2t^2_3)\right]^{5/2}},
\end{flalign}
where the same variable transformations and angular integrations were performed as in the case for the projection to the fermionic momenta $k_{x,y}k_z$. Subsequently, we use the ratios of fermionic and bosonic anisotropy parameters $\gamma_{21}=c_z\beta^2_{21}/c_p$ and $\gamma_{31}=c_p\beta^2_{31}/c_z$ as well as introduce the integration variable $r=q/Q$ to perform the integration over the variable $Q$
\begin{flalign}\label{eq:ProjectionFermionicMomentaK^2xy}
	\dfrac{\text{Tr}\left(\sigma_3\frac{\partial^2}{\partial k^2_{x,y}}\Sigma(\bold{k})\right)}{t^2_2\text{Tr}\left(\sigma_3\right)}|_{k_{x,y,z}=0}=-\alpha\,F_2(\gamma_{21},\gamma_{31})\,\text{ln}\ell,
\end{flalign}
where we have introduced the function
\begin{equation}
	F_2(\gamma_{21},\gamma_{31}) =\dfrac{3}{2\pi}\int^\infty_0rdr\,\dfrac{8\pi\left(-r^2\gamma_{21}+\gamma_{31}\right)(-2r^2+r^4\beta^2_{21}r^4+4r^2\gamma_{21\gamma_{31}+}\gamma^4_{31})}{\gamma_{21}\left(r^2+1\right)\left[4r^2+(r^2\gamma^2_{21}-\gamma^2_{31})^2\right]^{5/2}}.
\end{equation}
\subsubsection{Projection to $k^2_z$ fermionic momentum}
The projected fermionic self-energy to momentum $k^2_z$ reads
\begin{flalign}
	\dfrac{\text{Tr}\left(\sigma_3\frac{\partial^2\Sigma(\bold{k})}{\partial k^2_z}\right)}{t^2_3\text{Tr}\left(\sigma_3\right)}|_{k_{x,y,z}=0}=-\dfrac{e^2}{2(2\pi)^3}\int^\infty_0qdq\int^\Lambda_{\Lambda/\ell}Q^{d-1}dQ\,\dfrac{4 q^2t^4_1\left(-q^6t^6_2+5q^2Q^4t^2_2t^4_3+3Q^6t^6_3-q^4Q^2t^2_2(-8t^4_1+7t^2_2t^2_3)\right)}{\left(c^2_pq^2+c^2_zQ^2\right)\left[q^4t^4_2+Q^4t^4_3+2q^2Q^2(2t^4_1-t^2_2t^2_3)\right]^{5/2}},
\end{flalign}
where the same variable transformations and angular integrations were performed as in the case for the projection to $k_{x,y}k_z$. Subsequently, we use the ratios of fermionic and bosonic anisotropy parameters $\gamma_{21}=c_z\beta^2_{21}/c_p$ and $\gamma_{31}=c_p\beta^2_{31}/c_z$ as well as introduce the integration variable $r=q/Q$ to perform the integration over the variable $Q$
\begin{flalign}\label{eq:ProjectionFermionicMomentaK^2z}
	\dfrac{\text{Tr}\left(\sigma_3\frac{\partial^2}{\partial k^2_z}\Sigma(\bold{k})\right)}{t^2_3\text{Tr}\left(\sigma_3\right)}|_{k_{x,y,z}=0}=-\alpha\,F_3(\gamma_{21},\gamma_{31})\,\text{ln}\ell,
\end{flalign}
where we have introduced the function
\begin{equation}
	F_3(\gamma_{21},\gamma_{31}) =\dfrac{3}{2\pi}\int^\infty_0rdr\,\dfrac{8\pi r^2(8r^4\gamma_{21}-r^6\gamma^3_{21}-7r^4\gamma^2_{21}\gamma_{31}+5r^2\gamma_{21}\gamma^2_{31}+3\gamma^3_{31})}{\gamma_{31}\left(r^2+1\right)\left[4r^2+(r^2\gamma^2_{21}-\gamma^2_{31})^2\right]^{5/2}}.
\end{equation}
\subsubsection{Renormalization-group equations from the $\epsilon$-expansion}
Collecting the contributions from Eqs.~\eqref{eq:ProjectionFermionicMomentaKxyKz}, \eqref{eq:ProjectionFermionicMomentaK^2xy} and \eqref{eq:ProjectionFermionicMomentaK^2z}, the fermionic self-energy reads
\begin{flalign}
	\Sigma(\bold{k})=-\alpha\,\text{ln}\ell\,F_1(\gamma_{21},\gamma_{31}) 2t^2_1k_z(k_x\sigma_1+k_y\sigma_2)-\alpha\,\text{ln}\ell\,F_2(\gamma_{21},\gamma_{31})t^2_2(k^2_x+k^2_y)+\alpha\,\text{ln}\ell\,F_3(\gamma_{21},\gamma_{31})t^2_3k^2_z.
\end{flalign}
Then, the three-dimensional renormalized fermionic action writes
\begin{flalign}\label{eq:FermionicAction}
	S_f=T\sum_{i\Omega}\int\dfrac{d^3k}{(2\pi)^3}\psi^\dag(\bold{k},i\omega)\left[-i\omega\mathbb{1}_2+H_3(\bold{k})-\Sigma(\bold{k})\right]\psi(\bold{k},i\omega).
\end{flalign}
Considering the characteristic length scale $L$ and the definitions in Eq.~\eqref{eq:DimensionlessQuantities1}, Eq.~\eqref{eq:FermionicAction} is made self-similar to the bare fermionic action upon introducing the dimensionless quantities
\begin{flalign}\label{eq:DimensionlessQuantities2}
	\widetilde{\psi}^{(\dag)}=\psi^{(\dag)}L^{-\frac{7}{2}},\qquad\qquad\widetilde{k}=kL,\quad\qquad\widetilde{t}^2_{1,2,3}=t^2_{1,2,3}L^0.
\end{flalign}
The one-loop RG equations for the dimensionless fermionic anisotropy parameters $\widetilde{t}^2_{1,2,3}$ are
\begin{flalign}\label{eq:FermionicRGEquations}
	\dfrac{1}{\widetilde{t}^2_{1,2,3}}\dfrac{d\widetilde{t}^2_{1,2,3}}{d\text{ln}\ell}=\alpha\,F_{1,2,3}(\gamma_{21},\gamma_{31}).
\end{flalign}
\subsection{Phase portrait and consistency with large-$N_f$ analysis}
Since the RG equations for bosonic and fermionic anisotropy parameters in Eqs.~\eqref{eq:BosonicRGEquations} and \eqref{eq:FermionicRGEquations} are controlled by the parameters $\{\alpha,\gamma_{21},\gamma_{31}\}$, we construct the corresponding RG equations
 \begin{figure}[t]
	\centering
	\subfloat{\includegraphics[width=0.32\textwidth]{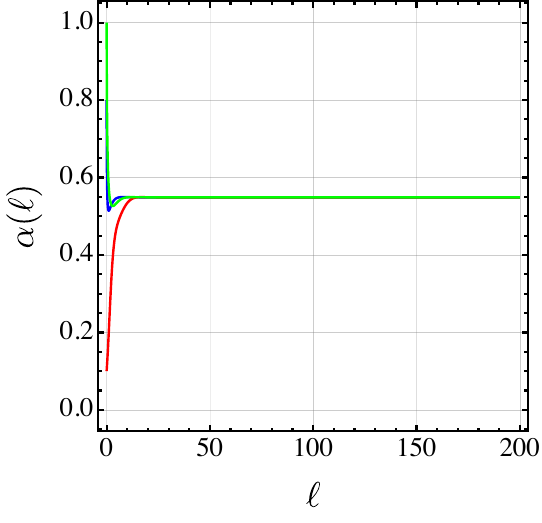}}\quad
	\subfloat{\includegraphics[width=0.32\textwidth]{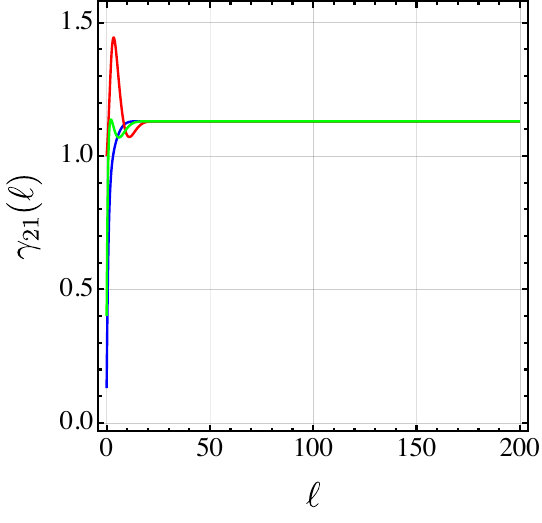}}\quad
	\subfloat{\includegraphics[width=0.32\textwidth]{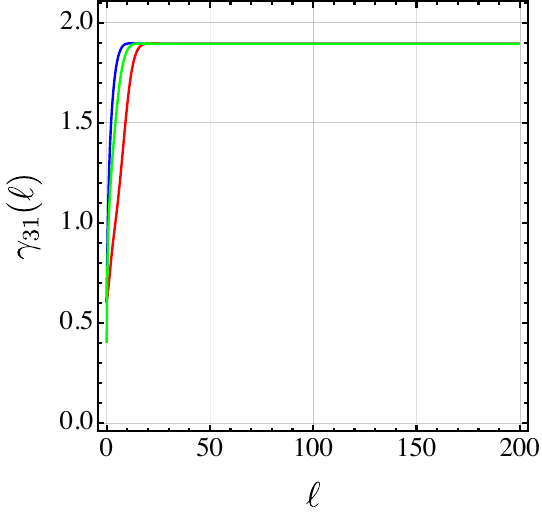}}
	\caption{Evolution of the RG parameters $(\alpha,\gamma_{21},\gamma_{31})$ according to Eq.~\eqref{eq:RGEquations} when $\epsilon=N_f=1$ in terms of the dimensionless scaling parameter $\ell$. The color code in the three plots represent the initial values for the aforementioned three parameters, $(\alpha^{(0)},\gamma^{(0)}_{21},\gamma^{(0)}_{31})$: $(0.8,0.13,0.6)$ (blue), $(0.1,1,0.6)$ (red), $(1,0.4,0.4)$ (green).}
	\label{fig:EvolutionRGParameters}
\end{figure}
\begin{flalign}\label{eq:RGEquations}
		\dfrac{1}{\gamma_{21,31}}\dfrac{d\gamma_{21,31}}{d\text{ln}\ell}=&\pm N_f\alpha\dfrac{B(\gamma_{31},\gamma_{21})-B(\gamma_{21},\gamma_{31})}{2}+\alpha\Big[F_{2,3}(\gamma_{21},\gamma_{31})-F_1(\gamma_{21},\gamma_{31})\Big],\nonumber\\
		\dfrac{1}{\alpha}\dfrac{d\alpha}{d\text{ln}\ell}=&\epsilon-N_f\alpha\dfrac{B(\gamma_{31},\gamma_{21})+B(\gamma_{21},\gamma_{31})}{2}-\alpha F_1(\gamma_{21},\gamma_{31}).
\end{flalign}
As a function of the dimensionless parameter $\ell$, we plot in Fig.~\ref{fig:EvolutionRGParameters} the evolution of the RG parameters $\{\alpha,\gamma_{21},\gamma_{31}\}$ for $\epsilon=N_f=1$. It is clear that the presence of Coulomb interaction leads to saturation values for the three RG parameters which indicate the \textit{existence} of an interacting fixed point. For the values $\epsilon=\{0.1,0.5,1\}$ when $N_f=1$, we find the fixed-point estimates
\begin{flalign}\label{eq:FixedPointSolutionsVarousEpsilon}
	    (\alpha^*,\gamma^*_{21},\gamma^*_{31})|^{N_f=1}_{\epsilon=0.1}&\approx(0.0549,1.129,1.894),\nonumber\\
	    (\alpha^*,\gamma^*_{21},\gamma^*_{31})|^{N_f=1}_{\epsilon=0.5}&\approx(0.2745,1.129,1.894),\nonumber\\
	    (\alpha^*,\gamma^*_{21},\gamma^*_{31})|^{N_f=1}_{\epsilon=1}&\approx(0.549,1.129,1.894).
\end{flalign}
We note that the fixed-point value for the coupling constant assumes the form $\alpha^*\propto\epsilon/2$ in agreement with the expectation that, on one hand, the renormalized coupling constant at one-loop order is proportional to the distance from the \textit{upper critical dimension} $(\epsilon=0)$, and on the other hand, that the fixed-point value is dominated by the renormalization of the bosonic spectrum i.e.~the functions $B(\gamma_{31},\gamma_{21})$ and $B(\gamma_{21},\gamma_{31})$ \cite{Abrikosov1974}. Furthermore, it is also noteworthy that the fixed-point values $\{\gamma^*_{21},\gamma^*_{31}\}$ are invariant with respect to parameter $\epsilon$ for \textit{fixed} $N_f$; this is due to the fact that the coupling constant $\alpha$ appears as a universal prefactor in the RG equations for the parameters $\{\gamma_{21},\gamma_{31}\}$ given in Eq.~\eqref{eq:RGEquations}. Regarding the bosonic anisotropy parameters $\{\widetilde{c}^2_p,\widetilde{c}^2_z\}$, the corresponding RG equations given in Eq.~\eqref{eq:BosonicRGEquations} \textit{at} the three-dimensional interacting fixed point read
\begin{equation}
	\left(\dfrac{1}{\widetilde{c}^2_p}\dfrac{d\widetilde{c}^2_p}{d\text{ln}\ell},\,\dfrac{1}{\widetilde{c}^2_z}\dfrac{d\widetilde{c}^2_z}{d\text{ln}\ell}\right)|^{N_f=1}_{\epsilon=1}\approx\left(0.599,1.006\right).
\end{equation}
The larger rate of divergence for the anisotropy parameter $\widetilde{c}^2_z$ compared to the one for $\widetilde{c}^2_p$ implies that the renormalized bosonic spectrum shows strong spatial anisotropy. Such a result is qualitatively consistent with the large-$N_f$ prediction given in Eq.~\eqref{eq:RenormalizedCoulombPotentialFixedPoint} about the renormalized Coulomb potential. On the other hand, considering the RG equations for the fermionic anisotropy parameters at the interacting fixed point in three spatial dimensions, we find
\begin{flalign}
	\left(\dfrac{1}{\widetilde{t}^2_1}\dfrac{d\widetilde{t}^2_1}{d\text{ln}\ell},\,\dfrac{1}{\widetilde{t}^2_2}\dfrac{d\widetilde{t}^2_2}{d\text{ln}\ell},\,\dfrac{1}{\widetilde{t}^2_3}\dfrac{d\widetilde{t}^2_3}{d\text{ln}\ell}\right)|^{N_f=1}_{\epsilon=1}=\left(0.197,-0.006,0.4\right).
\end{flalign}
The positiveness of the RG equations for the fermionic anisotropy parameters $\{t^2_1,t^2_2\}$ implies their divergence in the limit of infinite scale transformations, whereas the negativeness of the RG equation for $t^2_2$ leads the latter to vanish. That result concludes that the fermionic anisotropy parameter $t^2_2$, and in extend the term $(k^2_x+k^2_y)\sigma_z$ in the fermionic Hamiltonian given in Eq.~\eqref{eq:QuadraticHamiltonian}, is \textit{irrelevant} for the existence of the stable interacting fixed point, which qualitatively agrees with the conclusion derived within the large-$N_f$ analysis. For the ratios of fermionic anisotropy parameters $\beta_{21,31}=t_{2,3}/t_1$, the three-dimensional RG equations at the interacting fixed point write
\begin{equation}\label{eq:RatiosFermionicAnisotropiesAtTheInteractingFixedPoint}
	\left(\dfrac{1}{\beta^2_{21}}\dfrac{d\beta^2_{21}}{d\text{ln}\ell},\,\dfrac{1}{\beta^2_{31}}\dfrac{d\beta^2_{31}}{d\text{ln}\ell}\right)|^{N_f=1}_{\epsilon=1}\approx\left(-0.203,0.203\right).
\end{equation}
Thus, the ratio $\beta_{21}$ $(\beta_{31})$ negatively (positively) diverges as a function of the scale transformations and it tends towards zero (infinity), confirming the prediction in Eq.~\eqref{eq:LargeNfFixedPoint} derived using large-$N_f$ analysis. Such an outcome also agrees with the aforementioned result regarding the fermionic anisotropy parameter $t^2_2$ and it consolidates the argument about the emergence of spatial anisotropy in the presence of Coulomb interactions. Furthermore, the \textit{renormalized} dynamic critical exponent $\eta_\omega$ reads
\begin{equation}
	\eta_\omega=z-\alpha F_1(\gamma_{21},\gamma_{31}),
\end{equation}
while the fermionic and bosonic anomalous dimensions, $\eta_\psi$ and $\eta_\phi$ respectively, are given by the expressions
\begin{flalign}\label{eq:FermionicBosonicAnomalousDimensions}
	&\eta_\psi=\alpha F_1(\gamma_{21},\gamma_{31}),\nonumber\\
	&\eta_\phi=\dfrac{\alpha}{4}\left[2F_1(\gamma_{21},\gamma_{21})+N_fB(\gamma_{21},\gamma_{31})+N_fB(\gamma_{31},\gamma_{21})\right].
\end{flalign}
At the interacting fixed point in three spatial dimensions with $z=2$ and $N_f=1$, they assume the values
\begin{equation}\label{eq:CriticalExponents}
	\left(\eta_\omega,\eta_\psi,\eta_\phi\right)|^{N_f=1}_{\epsilon=1}=\left(1.803,0.197,0.5\right).
\end{equation}
The deviation of the dynamic critical exponent $\eta_\omega$ from its non-interacting value indicates anisotropic scaling among the spatial and temporal degrees of freedom in the fermionic Hamiltonian, or equivalently, among momentum and energy, i.e.~if momentum has engineering dimensions $L^{-2}$ for a generic quadratic semimetal, then the energy has engineering dimensions $L^{-\eta_\omega}$. The non-zero values of the bosonic and fermionic anomalous dimensions agree with the qualitative predictions derived using the large-$N_f$ method regarding, on one hand, the strong renormalization of the bosonic spectrum, and on the other, hand, the breakdown of the Fermi-liquid behavior. Quantitatively, the prediction for the bosonic anomalous dimension $\eta_\phi$ in Eq.~\eqref{eq:CriticalExponents} is identical to the estimated value from the large-$N_f$ analysis discussed in Eq.\eqref{eq:RenormalizedCoulombPotential}. Concerning the fermionic anomalous dimension $\eta_\psi$ given by the expression in Eq.~\eqref{eq:FermionicBosonicAnomalousDimensions}, we numerically identify the following scaling relation for values $N_f\geq100$ at three spatial dimensions:
\begin{flalign}
	\eta_\psi|^{N_f\geq100}_{\epsilon=1}\approx\dfrac{0.036}{N_f}.
\end{flalign}
The latter prediction agrees well to the result ${\eta^{(N_f\to\infty)}_\psi}^*\approx0.034/N_f$ obtained within the large-$N_f$ scheme.
\section{Physical observables in the interacting limit}
In the present section, we provide details regarding the calculation of the physical observables discussed in the relevant section in the main text. We determine the corresponding scaling relations in terms of the probed energy or temperature as well as the dependence of those physical observables on the three fermionic anisotropy parameters $\{t^2_1,t^2_2,t^2_3\}$ using the three-dimensional fermionic Hamiltonian given in Eq.~\eqref{eq:QuadraticHamiltonian}.
\subsection{Density of states}
The retarded fermionic Green's function reads
\begin{equation}
	G^R_3(\bold{k},\omega+i0^+)=\dfrac{P_+(\hat{\bold{k}})}{2}\dfrac{1}{\omega+i0^+-E_3(\bold{k})}+\dfrac{P_-(\hat{\bold{k}})}{2}\dfrac{1}{\omega+i0^++E_3(\bold{k})}
\end{equation}
where $P_{\pm}(\hat{\bold{k}})=\mathbb{1}_2\pm f(\hat{\bold{k}},\sigma_x,\sigma_y,\sigma_z)$ are projection operators to the conduction and valence bands with the function $f(\hat{\bold{k}},\sigma_x,\sigma_y,\sigma_z)$ depending linearly on the involved Pauli matrices. Therefore, the density of states at the interacting fixed point in three spatial dimensions is given by the expression \cite{FetterWalecka}
\begin{flalign}\label{eq:DensityOfStates}
	\rho(\omega)|^{N_f=1}_{\epsilon=1}&=-\dfrac{1}{\pi}\int\dfrac{d^3k}{(2\pi)^3}\text{Tr}\left[\text{Im}\,G^R_3(\bold{k},\omega+i0^+)\right]|^{N_f=1}_{\epsilon=1}\nonumber\\
	&\approx1.057\dfrac{\sqrt{|\omega|}}{4\pi^3t^2_2t_3}|^{N_f=1,\epsilon=1}_{t^*_2,\,t^*_3},
\end{flalign} 
where we used the corresponding result from Eq.~\eqref{eq:FixedPointSolutionsVarousEpsilon} to find
\begin{equation}\label{eq:BetaAtTheInteractingFixedPoint}
	\gamma^*_{21}\gamma^*_{31}|^{N_f=1}_{\epsilon=1}={\beta^2}^*|^{N_f=1}_{\epsilon=1}\approx2.1403.
\end{equation}
\subsection{Free energy}
Adding the chemical potential term $\mu$ to the fermionic Matsubara Green's function in Eq.~\eqref{eq:FermionicMatsubaraGreen'sFunction}, the free energy reads \cite{FetterWalecka}
\begin{flalign}
	F(T,\mu)=&-T\,\text{ln}\,\text{det}\left[-\dfrac{G^{-1}_3(\bold{k},i\omega)}{T}\right]\nonumber\\
	=&-T\int\dfrac{d^3k}{(2\pi)^3}\left\{\dfrac{E_3(\bold{k})}{T}+\text{ln}\left[1+\text{Exp}\left(-\dfrac{E_3(\bold{k})+\mu}{T}\right)\right]+\text{ln}\left[1+\text{Exp}\left(-\dfrac{E_3(\bold{k})-\mu}{T}\right)\right]\right\}.
\end{flalign}
Since the latter expression is divergent due to the presence of the term $\frac{E_3(\bold{k})}{T}$, we define the shifted free energy \cite{Wang2019}
\begin{equation}
	F_s(T,\mu)=F(T,\mu)+T\int\dfrac{d^3k}{(2\pi)^3}\dfrac{E_3(\bold{k})}{T}
\end{equation}
\subsubsection{Specific heat}
At the interacting fixed point in three spatial dimensions, the specific heat is given by the expression \cite{FetterWalecka}
\begin{flalign}\label{eq:SpecificHeat}
	C_v(T)|^{N_f=1}_{\epsilon=1}=-T\dfrac{\partial ^2F_s(T,\mu=0)}{\partial T^2}|^{N_f=1}_{\epsilon=1}\approx0.517\dfrac{15T^{3/2}}{4\pi^2t^2_2t_3}|^{N_f=1,\,\epsilon=1}_{t^*_2,\,t^*_3},
\end{flalign}
where Eq.~\eqref{eq:BetaAtTheInteractingFixedPoint} has been employed.
\subsubsection{Compressibility}
The compressibility at the interacting fixed point in three spatial dimensions writes \cite{FetterWalecka}
\begin{flalign}\label{eq:Compressbility}
	\kappa(T)|^{N_f=1}_{\epsilon=1}=\dfrac{\partial ^2F_s(T,\mu)}{\partial \mu^2}|^{N_f=1,\epsilon=1}_{\mu=0}\approx0.721\dfrac{T^{1/2}}{2\pi^2t^2_2t_3}|^{N_f=1,\,\epsilon=1}_{t^*_2,\,t^*_3},
\end{flalign}
where Eq.~\eqref{eq:BetaAtTheInteractingFixedPoint} has been employed.
\subsection{Diamagnetic susceptibility}
Introducing the current-density parameters
\begin{flalign}
	&J_x(\bold{k})=2t^2_1k_z\sigma_x+2t^2_2k_x\sigma_z,\nonumber\\
	&J_y(\bold{k})=2t^2_1k_z\sigma_y+2t^2_2k_y\sigma_z,\nonumber\\
	&J_z(\bold{k})=2t^2_1k_x\sigma_x+2t^2_1k_y\sigma_y-2t^2_3k_z\sigma_z,\nonumber\\
\end{flalign}
the magnetic susceptibility along the $i$-th spatial axis is given by the Fukuyama formula \cite{Fukuyama1970}
\begin{equation}
	\chi_i(T)=\sum_{m,n=\{x,y,z\}}\epsilon_{imn}T\sum_{i\omega}\int\dfrac{d^3k}{(2\pi)^3}\text{Tr}\left[G_3(\bold{k},i\omega)J_m(\bold{k})G_3(\bold{k},i\omega)J_n(\bold{k})G_3(\bold{k},i\omega)J_m(\bold{k})G_3(\bold{k},i\omega)J_n(\bold{k})\right],
\end{equation}
where $\epsilon_{imn}$ is the Levi-Civita symbol. 
Due to the $C^z_4$ rotational symmetry around the $\hat{z}$ axis, the $x$- and $y$-components of magnetic susceptibility at the three-dimensional interacting fixed point are equal and they read
\begin{flalign}\label{eq:DiamagneticSusceptibilityXXYY}
	\chi_x(T)|^{N_f=1}_{\epsilon=1}=\chi_y(T)|^{N_f=1}_{\epsilon=1}\approx-23.694\,e^2\sqrt{T}\,t_3|^{N_f=1,\,\epsilon=1}_{t^*_3},
\end{flalign}
whereas the $z$-component of the magnetic susceptibility writes
\begin{flalign}\label{eq:DiamagneticSusceptibilityZZ}
	\chi_z(T)|^{N_f=1}_{\epsilon=1}\approx-4.686\,e^2\sqrt{T}\dfrac{t^4_1}{t^3_3}|^{N_f=1,\,\epsilon=1}_{t^*_1,\,t^*_3}.
\end{flalign}
In both computations, Eq.~\eqref{eq:BetaAtTheInteractingFixedPoint} has been used.
\subsection{Optical conducitivity}
The current-current correlation function is given by the expression \cite{FetterWalecka}
\begin{equation}
	\Pi_{n,m}(i\Omega,T)=e^2T\sum_{i\omega}\int\dfrac{d^3k}{(2\pi)^3}\text{Tr}\left[J_m(\bold{k})G_3(\bold{k},i\omega)J_n(\bold{k})G_3(\bold{k},i\omega+i\Omega)\right],
\end{equation}
where the indices $n,m=\{x,y,z\}$ run over the spatial coordinates. The linear conductivity then reads
\begin{equation}
	\sigma_{m,n}(\Omega,T)=-\dfrac{\text{Im}\,\Pi_{m,n}(i\Omega,T)}{\Omega}.
\end{equation}
Due to the rotational $C^z_4$ symmetry around the $\hat{z}$ axis, we find the longitudinal conductivity components along the $\hat{x}$ and $\hat{y}$ coordination axes at the three-dimensional interacting fixed point to be equal reading
\begin{flalign}\label{eq:OpticalConductivityXXYY}
	&\sigma_{xx}(\Omega\to0,T)|^{N_f=1}_{\epsilon=1}=\sigma_{yy}(\Omega\to0,T)|^{N_f=1}_{\epsilon=1}\approx0.0976\,e^2\dfrac{T^{3/2}}{t_3}|^{N_f=1,\,\epsilon=1}_{t^*_3},\nonumber\\
	&\sigma_{xx}(\Omega\neq0,T)|^{N_f=1}_{\epsilon=1}=\sigma_{yy}(\Omega\neq0,T)|^{N_f=1}_{\epsilon=1}\approx0.00848\,e^2\dfrac{\sqrt{|\Omega|}}{t_3}\tanh\left(\dfrac{\Omega}{4T}\right)|^{N_f=1,\,\epsilon=1}_{t^*_3},
\end{flalign}
where Eq.~\eqref{eq:BetaAtTheInteractingFixedPoint} has been used. On the other hand, the longitudinal conductivity along the $\hat{z}$ coordination axis reads
\begin{flalign}\label{eq:OpticalConductivityZZ}
	&\sigma_{zz}(\Omega\to0,T)|^{N_f=1}_{\epsilon=1}\approx 0.0647e^2\dfrac{T^{3/2}t^3_3}{\pi^3t^4_1}|^{N_f=1,\,\epsilon=1}_{t^*_1,t^*_3},\nonumber\\
	&\sigma_{zz}(\Omega\neq0,T)|^{N_f=1}_{\epsilon=1}\approx0.00562 e^2\dfrac{\sqrt{|\Omega|}\,t^3_3}{4\sqrt{2}\pi^3t^4_1}\tanh\left(\dfrac{\Omega}{4T}\right)|^{N_f=1,\,\epsilon=1}_{t^*_1,t^*_3},
\end{flalign}
\subsection{Scaling relations of the physical observables in the interacting limit}
First, we write the RG equations for the fermionic anisotropy parameters $\{t^2_1,t^2_2,t^2_3\}$ \textit{at} the interacting fixed point:
\begin{flalign}\label{eq:RGEquationsFermionicAnisotropyParametersInteractingFixedPoint}
	&\dfrac{1}{t^2_1}\dfrac{dt^2_1}{d\text{ln}\ell}|^{N_f=1}_{\epsilon=1}=z^*-f^*_{x,y}-1+\alpha^*F_1(\gamma^*_{21},\gamma^*_{31})|^{N_f=1}_{\epsilon=1}\stackrel{!}{=}0,\nonumber\\
	&\dfrac{1}{t^2_2}\dfrac{dt^2_2}{d\text{ln}\ell}|^{N_f=1}_{\epsilon=1}=z^*-f^*_{x+y}+\alpha^*F_2(\gamma^*_{21},\gamma^*_{31})|^{N_f=1}_{\epsilon=1}\stackrel{!}{=}0,\nonumber\\
	&\dfrac{1}{t^2_3}\dfrac{dt^2_3}{d\text{ln}\ell}|^{N_f=1}_{\epsilon=1}=z^*-\alpha^*F_3(\gamma^*_{21},\gamma^*_{31})|^{N_f=1}_{\epsilon=1}\stackrel{!}{=}0.
\end{flalign}
Due to the emerging spatial anisotropy at the interacting fixed point, already identified within both the large-$N_f$ analysis and the $\epsilon$-expansion, we introduced a different scaling for the individual coordination axes $\hat{x}$ and $\hat{y}$ compared to the norm in $(x,y)$-plane: $[k_{x,y}]=L^{-f_{x,y}}$ and $[k^2_x+k^2_y]=L^{-f_{x+y}}$. If the parameters $\{t^2_1,t^2_2,t^2_3\}$ were considered marginal, then one would set $f_{x+y}=2f_{x,y}=z=2$. Solving Eq.~\eqref{eq:RGEquationsFermionicAnisotropyParametersInteractingFixedPoint} in terms of the fixed-point values $\{z^*,f^*_{x,y},f^*_{x+y}\}$, we find
\begin{flalign}\label{eq:FixedPointSolutionsAtTheInteractingFixedPoint}
	&z^*=2-\alpha^*F_3(\gamma^*_{21},\gamma^*_{31})|^{N_f=1}_{\epsilon=1},\nonumber\\
	&f^*_{x,y}=2-\alpha^*\Big(F_3(\gamma^*_{21},\gamma^*_{31})-F_2(\gamma^*_{21},\gamma^*_{31})\Big)|^{N_f=1}_{\epsilon=1},\nonumber\\
	&f^*_{x+y}=1-\alpha^*\Big(F_3(\gamma^*_{21},\gamma^*_{31})-F_1(\gamma^*_{21},\gamma^*_{31})\Big)|^{N_f=1}_{\epsilon=1}.
\end{flalign}
To compute the scaling relations of the physical observables in the presence of the scale-depedent fermionic anisotropy parameters $\{t^2_1,t^2_2,t^2_3\}$, we use the following relation connecting the bare quantity $p_0$ and the renormalized quantity $p_R$ under a \textit{single} (not necessarily infinitesimal) scale transformation \cite{Goldenfeld2018}
\begin{equation}\label{eq:CorrespondenceBetweenRenormalizedAndBareQuantitites}
	p_R=\ell^{d(p_0)}p_0.
\end{equation}
The dimensionless scaling parameter is $\ell$ and $d(p_0)$ is an exponent associated to the engineering dimensions of the bare quantity $p_0$ in terms of a characteristic length $L$. Considering the engineering dimensions $[\nu_0]=[T_0]=L^{-z}$ for an energy $\nu$ and temperature $T$ scale, we write $d(\nu_0)=d(T_0)=-z$ leading to the result
\begin{flalign}\label{eq:Cutoffs}
	\ell=\left(\dfrac{\nu_0}{\nu_R}\right)^{1/z}=\left(\dfrac{T_0}{T_R}\right)^{1/z}.
\end{flalign}
Then, one can interpret the quantities $\nu_0$ $(T_0)$ and $\nu_R$ $(T_R)$ as the UV and IR energy (temperature) cutoffs, respectively. Using the correspondence in Eq.~\eqref{eq:CorrespondenceBetweenRenormalizedAndBareQuantitites} for the expressions given in Eqs.~\eqref{eq:DensityOfStates}, \eqref{eq:Compressbility}, \eqref{eq:SpecificHeat}, \eqref{eq:DiamagneticSusceptibilityXXYY}, \eqref{eq:DiamagneticSusceptibilityZZ}, \eqref{eq:OpticalConductivityXXYY} and \eqref{eq:OpticalConductivityZZ} we find that
\begin{flalign}
	&\rho(\nu_R)|^{N_f=1}_{\epsilon=1}=\left(\dfrac{\nu_0}{\nu_R}\right)^{\dfrac{-\frac{z^*}{2}-(1+f^*_{x+y}-z^*+\frac{2-z^*}{2})}{z^*}}\rho(\nu_0),\nonumber\\
	&C_v(T_R)|^{N_f=1}_{\epsilon=1}=\left(\dfrac{T_0}{T_R}\right)^{\dfrac{-\frac{3z^*}{2}-(1+f^*_{x+y}-z^*+\frac{2-z^*}{2})}{z^*}}C_v(T_0),\nonumber\\
	&\kappa(T_R)|^{N_f=1}_{\epsilon=1}=\left(\dfrac{T_0}{T_R}\right)^{\dfrac{-\frac{z^*}{2}-(1+f^*_{x+y}-z^*+\frac{2-z^*}{2})}{z^*}}\kappa(T_0),\nonumber\\
	&\chi_{x,y}(T_R)|^{N_f=1}_{\epsilon=1}=\left(\dfrac{T_0}{T_R}\right)^{\dfrac{-\frac{z^*}{2}+\frac{2-z^*}{2}}{z^*}}\chi_{x,y}(T_0),\nonumber\\
	&\chi_z(T_R)|^{N_f=1}_{\epsilon=1}=\left(\dfrac{T_0}{T_R}\right)^{\dfrac{-\frac{z^*}{2}+4\frac{1+f^*_{x,y}-z^*}{2}-3\frac{2-z^*}{2}}{z^*}}\chi_{x,y}(T_0),\nonumber\\
	&\sigma^\text{DC}_{xx,yy}(T_R)|^{N_f=1}_{\epsilon=1}\equiv\sigma_{xx,yy}(\nu\to0,T_R)|^{N_f=1}_{\epsilon=1}=\left(\dfrac{T_0}{T_R}\right)^{\dfrac{-\frac{3z^*}{2}+\frac{2-z^*}{2}}{z^*}}\sigma_{xx,yy}(T_0),\nonumber\\
	&\sigma^{T\to\infty}_{xx,yy}(\nu_R)|^{N_f=1}_{\epsilon=1}\equiv\sigma_{xx,yy}(\nu_R,T\to\infty)|^{N_f=1}_{\epsilon=1}=\left(\dfrac{\nu_0}{\nu_R}\right)^{\dfrac{-\frac{z^*}{2}-\frac{2-z^*}{2}}{z^*}}\sigma_{xx,yy}(\nu_0,T\to\infty)\nonumber\\
	&\sigma^\text{DC}_{zz}(T_R)|^{N_f=1}_{\epsilon=1}\equiv\sigma_{zz}(\nu\to0,T_R)|^{N_f=1}_{\epsilon=1}=\left(\dfrac{T_0}{T_R}\right)^{\dfrac{-\frac{3z^*}{2}+3\frac{2-z^*}{2}-4\frac{1+f^*_{x,y}-z^*}{2}}{z^*}}\sigma_{zz}(T_0),\nonumber\\
	&\sigma^{T\to\infty}_{zz}(\nu_R)|^{N_f=1}_{\epsilon=1}\equiv\sigma_{zz}(\nu_R,T\to\infty)|^{N_f=1}_{\epsilon=1}=\left(\dfrac{\nu_0}{\nu_R}\right)^{\dfrac{-\frac{z^*}{2}+3\frac{2-z^*}{2}-4\frac{1+f^*_{x,y}-z^*}{2}}{z^*}}\sigma_{zz}(\nu_0,T\to\infty).
\end{flalign} 
Using Eq.~\eqref{eq:FixedPointSolutionsAtTheInteractingFixedPoint} in conjuction with the fixed-point values for the parameters $\{\alpha,\gamma_{21},\gamma_{31}\}$ at $\epsilon=1$ given in Eq.~\eqref{eq:FixedPointSolutionsVarousEpsilon}, we arrive at the following scaling relations in terms of the probed IR energy or temperature scale
\begin{flalign}
	&\rho(\nu_R)|^{N_f=1}_{\epsilon=1}\propto\nu^{\frac{1}{2}+\eta_1}_R,\qquad C_v(T_R)|^{N_f=1}_{\epsilon=1}\propto T^{\frac{3}{2}+\eta_1}_R,\qquad\kappa(T_R)|^{N_f=1}_{\epsilon=1}\propto T^{\frac{1}{2}+\eta_1}_R,\nonumber\\
	&\chi_{x,y}(T_R)|^{N_f=1}_{\epsilon=1}\propto T^{\frac{1}{2}-\eta_2}_R,\qquad\chi_z(T_R)|^{N_f=1}_{\epsilon=1}\propto T^{\frac{1}{2}+\eta_3}_R,\qquad\sigma^\text{DC}_{xx,yy}(T_R)|^{N_f=1}_{\epsilon=1}\propto T^{\frac{3}{2}+\eta_2}_R,\nonumber\\
	&\sigma^{T\to\infty}_{xx,yy}(\nu_R)|^{N_f=1}_{\epsilon=1}\propto\nu^{\frac{1}{2}+\eta_2}_R,\qquad\sigma^\text{DC}_{xx,yy}(T_R)|^{N_f=1}_{\epsilon=1}\propto T^{\frac{3}{2}-\eta_3}_R,\qquad\sigma^{T\to\infty}_{xx,yy}(\nu_R)|^{N_f=1}_{\epsilon=1}\propto\nu^{\frac{1}{2}-\eta_3}_R,
\end{flalign}
where the corrections read $\eta_1\approx0.121$, $\eta_2\approx0.125$ and $\eta_3\approx0.129$. The observables $\{\rho(\omega),C_v(T),\kappa(T)\}$ are identically suppressed due to the reduction of the free energy. On the other hand, long-range Coulomb interactions enhance the diamagnetic response perpendicular to the rotational $\hat{z}$ axis, because the emerging spatial anisotropy tends to develop a finite magnetic moment in the $(x,y)$-plane that minimizes the total magnetic energy \cite{Ashcroft1976}. Similarly, the enhancement of the optical conductivity along the rotational $\hat{z}$ axis can be explained due to the largely screened Coulomb potential along that axis (see Eq.~\eqref{eq:RenormalizedCoulombPotential}) leading to the broadening of bandwidth in the conduction band of fermions.
\subsection{Flux of Berry curvature in the interacting limit}
The flux of Berry curvature through a half-plane is defined as a two-dimensional integral over a surface of constant energy \cite{Zhuang2024}
\begin{equation}
	C^{\pm}=\pm\dfrac{1}{2\pi}\int\int\bold{\Omega}(\bold{k})d\bold{S}(\bold{k})|_{E_3(\bold{k})=\text{const.},\,\text{half plane}}.
\end{equation}
The signs $(+)$ and $(-)$ denote the contribution from the conduction and valence band, respectively. Parameter $E_3(\bold{k})$ is the three-dimensional dispersion relation, and $\bold{\Omega}(\bold{k})$ is the vector of Berry curvature reading
\begin{flalign}
	\Omega_n(\bold{k})=\epsilon_{ijn}\dfrac{\bold{d}(\bold{k})\cdot\Big(\partial_{k_i}\bold{d}(\bold{k})\times\partial_{k_j}\bold{d}(\bold{k})\Big)}{2E^3_3(\bold{k})},
\end{flalign}
with the vector of three-dimensional form factors writing $\bold{d}(\bold{k})=\Big(d_1(\bold{k}),d_2(\bold{k}),d_3(\bold{k})\Big)$. For the fermionic Hamiltonian given in Eq.~\eqref{eq:QuadraticHamiltonian} with \textit{scale-dependent} fermionic anisotropy parameters $\{t^2_1,t^2_2,t^2_3\}$, we arrive at the following expressions for the fluxes of Berry curvature through the upper- and lower-half $\hat{z}$-planes
\begin{flalign}\label{eq:FluxBerryCurvature}
	C^{\pm}_{z>0}=\pm\dfrac{\beta^2_{21}+\beta^2_{31}}{2},\qquad\qquad	C^{\mp}_{z<0}=\mp\dfrac{\beta^2_{21}+\beta^2_{31}}{2}.
\end{flalign}
If the fermionic anisotropy parameters were considered scale-independent, they \textit{should} satisfy the relation $t_1=\sqrt{t_2t_3}$ due to the Hopf map \cite{Moore2008}. The fluxes of Berry curvature would then be quantized leading the fermionic Hamiltonian in Eq.~\eqref{eq:QuadraticHamiltonian} to correspond to a Berry-dipole semimetal \cite{Zhuang2024}. However, the presence of long-range Coulomb interactions introduces scale-dependent fermionic anisotropy parameters that \textit{break} the quantization of the fluxes of Berry curvature, since the relation $t_1=\sqrt{t_2t_3}$ is no longer satisfied. As a result, considering Eq.~\eqref{eq:RatiosFermionicAnisotropiesAtTheInteractingFixedPoint}, the fluxes of Berry curvature become \textit{infinite} at the three-dimensional interacting fixed point:
\begin{flalign}
	C^{\pm}_{z>0}|^{N_f=1}_{\epsilon=1}\to\pm\infty,\qquad\qquad	C^{\mp}_{z<0}|^{N_f=1}_{\epsilon=1}\to\mp\infty.
\end{flalign}

\bibliography{NFLBerryDipole_SupplementalMaterial}

\clearpage